\documentclass[nonacm, manuscript, screen]{acmart}

\usepackage{subcaption}
\usepackage{pifont}
\usepackage{mathtools}
\newcommand{\greenmark}{\textcolor{green}{\ding{52}}}
\newcommand{\orangemark}{\textcolor{orange}{\ding{106}}}
\newcommand{\greenCircMark}{$\mathrlap{\bigcirc}\mkern2mu$\greenmark}
\newcommand{\orangeCircMark}{$\mathrlap{\bigcirc}\mkern1mu$\orangemark}
\newcommand{\CircQuestionmark}{$\mathrlap{\bigcirc}\mkern4.5mu$?}

\newcommand{\filter}{\textit{Conversation Filter}}
\newcommand{\save}{\textit{Save-For-Later}}
\newcommand{\share}{\textit{Peer Sharing}}
\newcommand{\studentFeedback}{\textit{Student Messaging}}
\newcommand{\tutorFeedback}{\textit{Feedback Inbox}}
\newcommand{\AIbehavior}{\textit{AI Behaviour Change}}

\AtBeginDocument{%
  }

\begin{document}


\title[What UI Design Can Do For ``Good'' Oversight Work]{What UI Design Can Do For ``Good'' Oversight Work: A Three-Month Action Research Study} 


\author{Cedric Faas}
\affiliation{%
  \institution{Saarland University, Saarland Informatics Campus}
  \city{Saarbrücken}
  \country{Germany}}
\email{faas@cs.uni-saarland.de}
\orcid{0009-0000-7918-4233}

\author{Lisa Dargasz}
\authornote{Authors contributed equally to this research.}
\email{lisa.dargasz@dfki.de}
\orcid{0009-0008-1642-5480}
\author{Julius Gabelmann}
\authornotemark[1]
\email{julius.gabelmann@dfki.de}
\orcid{0009-0001-6780-5685}
\author{Felix Jahn}
\authornotemark[1]
\email{felix.jahn@dfki.de}
\orcid{0000-0003-4851-3385}
\affiliation{%
  \institution{German Research Center for Artificial Intelligence (DFKI)}
  \city{Saarbrücken}
  \country{Germany}
}

\author{Markus Langer}
\affiliation{%
  \institution{Department of Psychology, University of Freiburg}
  \city{Freiburg im Breisgau}
  \country{Germany}}
\email{markus.langer@psychologie.uni-freiburg.de}
\orcid{0000-0002-8165-1803}

\author{Kevin Baum}
\affiliation{%
  \institution{Center for European Research in Trusted AI (CERTAIN), German Research Center for Artificial Intelligence (DFKI)}
  \city{Saarbrücken}
  \country{Germany}}
\email{kevin.baum@dfki.de}
\orcid{0000-0002-6893-573X}

\author{Anna Maria Feit}
\affiliation{%
  \institution{Saarland University, Saarland Informatics Campus}
  \city{Saarbrücken}
  \country{Germany}}
\email{feit@cs.uni-saarland.de}
\orcid{0000-0003-4168-6099}

\renewcommand{\shortauthors}{Faas, Dargasz, Gabelmann, Jahn, Langer, Baum, and Feit}

\begin{abstract}
With AI automating consequential tasks, human oversight is required for ethical, legal, and practical reasons. Yet these oversight tasks can be demanding and often make for poor work.
Empirical evidence on overseers’ lived work experiences remains underexplored, yet critical for designing  ``good’’ oversight.
We provide longitudinal, in-the-wild insights through action research accompanying (N=8) tutors who oversaw an AI-supported learning assistant over three months. Our thematic analysis combines insights from twelve weekly meetings and two workshop iterations and was enriched by four repeated questionnaires. 
We identified stimulation, mastery, and relatedness as key challenges, stemming from the task's repetitiveness, unclear impact of interventions, and limited connections with students and peers. Informed by our initial findings, we deployed targeted UI modifications with mixed success.
This paper documents the long-term evolution of the work characteristics of a real-world oversight task and discusses which oversight work characteristics can and cannot be supported by interface design.    
\end{abstract}




\maketitle

\section{Introduction}

As AI systems are increasingly involved in consequential tasks, the risks arising from their malfunctions, errors, (un)conscious misuse and undesirable outputs grow with them. One important layer of mitigating these risks is human oversight. Broadly understood, human oversight is a deliberate and often explicitly assigned activity in which one or more persons monitor the operation and outputs of an  AI system and intervene when professional judgment, established criteria, or legal provisions indicate that action is required to mitigate 
risks~\cite{gaube_keeping_2026}. 
The reasons for requiring such oversight are ethical~\cite{jobin_global_2019,crootof_humans_2023}, legal~\cite{laux_institutionalised_2023,green_flaws_2022}, and practical~\cite{sterz_quest_2024, green_flaws_2022}. However, oversight that fails misallocates responsibility and resources and risks undermining trust in the very systems it is meant to safeguard. Educational AI tools illustrate all three: they promise accessible, always available, and highly scalable support for students, but they may hallucinate, spread misinformation, or interact in pedagogically unproductive or even counterproductive ways. Oversight is one of the few mechanisms by which a duty of care towards students can be fulfilled once the teaching relationship is mediated by a system and may even be legally required, for instance, under the EU AI Act \cite[Art.~14]{euaiact2024}, when AI is involved in grading or assessing students.

Whether human oversight achieves these goals, however, is not settled by mandating it. Oversight and its effectiveness depend on different requirements: on what overseers can know about the system, on what they can causally affect, on whether they can hold their attention on the task and act on what they see, and on what motivates them to exercise their role~\cite{sterz_quest_2024}.  
Where these conditions are not met, overseers may be relegated to rubber-stamping AI decisions~\cite{green_flaws_2022, faas_ai_2026}, feel useless when highly accurate systems leave them little to contribute~\cite{vaccaro_when_2024,green_principles_2019}, and end up bearing responsibility for outcomes they could not effectively influence~\cite{elish_moral_2019, faas_dont_2026}. The last two points in particular are easy to overlook and difficult to meet, especially in the long term. Emerging research points to deteriorating psychological effects (e.g. mastery and autonomy) of oversight work~\cite{ehsan_future_2026, dhanorkar_human_2026,faas_ai_2026, mitchell_ai_2026} and stresses the need for socio-technical solutions that enable \textit{effective} human oversight~\cite{sterz_quest_2024}. Effectiveness is therefore not separable from the quality of the work. Designing for effective oversight means designing oversight work that meets overseers' psychological needs.
Drawing on work design literature~\cite{parker_smart_2024}, \citet{faas_design_2026} propose five characteristics of ``good'' oversight work: \textbf{\textit{Stimulating}}, \textbf{\textit{Mastery}}, \textbf{\textit{Autonomous}}, \textbf{\textit{Relational}}, \textbf{\textit{Tolerable}} (SMART). On this view, poor work design is not merely a burden on the oversight personnel---it is itself a failure mode of oversight.

Understanding the psychological challenges that oversight poses, therefore, requires examining all SMART characteristics, their interrelations, and how negative aspects can be mitigated. Moreover, oversight is typically performed alongside other work tasks and over prolonged periods, so its work characteristics must be studied in real-world settings and over time. Yet most empirical research examines only a subset of characteristics, in short-term lab studies, and outside participants' everyday work~\cite[e.g.][]{faas_ai_2026, ranjit_are_2026, reicherts_ai_2025}. This paper, therefore, addresses two questions: (RQ1) Which work characteristics does a human oversight task exhibit in practice, and how do they develop over an extended period of everyday work? (RQ2) Which of these characteristics can interface design address---and which resist it?

To answer these questions, we conducted one cycle of action research integrating in-the-wild observation with requirements-elicitation workshops and work-design theory. Over a three-month lecture period, we accompanied the tutors (N=8) of a university statistics lecture in overseeing an AI-supported learning assistant (ALA) that interacted with 159 students as part of their everyday tutoring job. Through twelve weekly meetings, two iterations of structured workshops, and four repeated questionnaires, we identified key challenges tutors experienced. Grounded in the design considerations proposed by \citet{faas_design_2026}, we designed six UI modifications with the goal of affecting all SMART factors. In the last weeks, tutors used the modified interface, allowing us to observe how interface design influenced the oversight work characteristics and the tutors' perception of their task.

In our thematic analysis, \textbf{\textit{stimulation}}, \textbf{\textit{mastery}}, and \textbf{\textit{relatedness}} emerged as the main challenges of the oversight task, while it was perceived as mostly \textit{autonomous} and \textit{tolerable}. Most prominently, tutors disliked the repetitiveness of the task, the indirect and invisible impact of their interventions, and the lack of connection to students and fellow tutors. Our UI modifications addressed these challenges with mixed success: the \filter\ feature let tutors deliberately focus their effort and made the task more stimulating, and the \share\ feature created a valued sense of community among peers---one that tutors even wished to extend toward students. Yet the tutors' uncertainty about whether their interventions mattered persisted. This, we argue, points to a limit of what interface design can compensate: where overseers lack a closed loop from their interventions to the system's behaviour and to those affected by it, the missing sense of impact is not an interface problem but an oversight problem. 

As one cycle of action research built on longitudinal field observation, our findings are specific to the task and tutors studied, and we discuss their transferability rather than claim generality. We expect oversight tasks of this shape---asynchronous review of large volumes of AI interactions---to become increasingly common as ever more autonomous and agentic systems demand oversight that scales~\cite{gaube_keeping_2026, zhu_designing_2026}. Against this background, we conjecture that repetitiveness and the resulting lack of stimulation will be a recurring concern that manifests differently across oversight tasks, that challenges to mastery will arise wherever the effects of interventions remain invisible to overseers, and that limited relatedness deserves deliberate attention in the design of oversight work and interfaces. Our study provides evidence that the work characteristics of human oversight can be supported not only through task and work design, but also through interface and interaction design, and it delineates where interface design reaches its limits.
\section{Background and Related Work}
\label{sec:theoreticalBackground}

\subsection{Human Oversight of AI }

Across domains, AI systems are getting increasingly capable of performing tasks autonomously~\cite[see][for an overview]{lai_towards_2023, hosseini_role_2025}. While ethical~\cite{jobin_global_2019, crootof_humans_2023} and legal~\cite{laux_institutionalised_2023, green_flaws_2022} reasons to oversee these systems mostly hold in high-risk contexts, there are also practical reasons to do so: Humans can monitor the AI interactions to identify malfunctions, inaccurate, inappropriate, or in other ways undesirable AI behaviour, and intervene to address risks. While most prior work focused on human oversight in high-risk tasks, we expect overseeing increasingly agentic AI systems in less critical tasks---as we investigate here---to become more prominent among a wider range of stakeholders.
Human oversight is broadly understood as the supervision of an AI system by at least one natural person who typically holds the authority to influence the system's operations or effects~\cite{sterz_quest_2024}. More concretely, it has been characterised as a deliberate, explicitly assigned activity in which one or more persons monitor the operation and output of an AI system and intervene when professional judgment, established criteria, or legal provisions indicate that action is required to mitigate risks~\cite{gaube_keeping_2026}.

Human oversight can be realised in structurally different ways. Following the causal taxonomy of \citet{baum_constitutive_2026}, being \textit{in} the loop and being \textit{on} the loop differ in kind rather than in degree: a human-in-the-loop (HITL) is \textit{constitutive}---their contribution is necessary for the process to generate an output at all---whereas a human-on-the-loop (HOTL) is \textit{corrective}: external to the default causal chain, but able to prevent, modify, or override what that chain produces. 
On this account, human oversight in the normative sense is a mode of HOTL rather than a mode of HITL. A human who must approve every output is a gate in the process, and a gate passed as a matter of routine---the familiar rubber-stamping failure~\cite{green_flaws_2022}---supplies no oversight at all. 
A second, orthogonal dimension of the taxonomy concerns cognitive integration: whether overseer and system remain \emph{complementary}---each forming its judgment independently---or become \emph{hybrid}---the overseer's task-relevant judgments are partly constituted by the system's representations~\cite{baum_constitutive_2026}. The distinction matters for oversight because effective assessment presupposes a capacity for judgment at least partly independent of the system under scrutiny, which representational interdependence erodes.

Within HOTL, three temporal modes are further distinguished: \emph{synchronous}, \emph{asynchronous}, and \emph{anticipatory} oversight. In synchronous oversight, humans monitor operations in real time and can intervene immediately. In asynchronous oversight, humans review system operations periodically, e.g., in logs or audits. In anticipatory oversight, humans shape the space of permissible actions pre-deployment or between cycles in which the system operates~\cite{baum_constitutive_2026}. This mode connects to work framing personalised, normative feedback-driven alignment as an application context for explainability, where the justifiability and verification of a system's ``understanding'' of normative expectations become central~\cite{baum_principals_2026}.

In our study setting, tutors review chat protocols of student-AI interactions and provide structured feedback on the AI's answers. Within this taxonomy, our study setting constitutes \emph{asynchronous} HOTL: the system is reviewed periodically, and the corrective capacity is temporally displaced~\cite{baum_constitutive_2026}. Since this temporal displacement allows for the greater scaling required by ever more autonomous and agentic systems, we expect oversight tasks of this shape to become increasingly common~\cite{gaube_keeping_2026, zhu_designing_2026}.

\subsection{Work Characteristics of Human Oversight}
Prior work on human oversight and human-AI interaction has mostly focused on outcomes rather than the people involved; e.g., XAI research has aimed to boost joint performance with mixed success~\cite{longo_explainable_2024, ali_explainable_2023, vaccaro_when_2024}. Building on the long-standing role of psychological processes in interface design~\cite{horvitz_principles_1999,bellotti_intelligibility_2001, friedman_value-sensitive_1996}, recent work advocates more human-centered perspectives~\cite{sterz_quest_2024, faas_design_2026, shneiderman_human-centered_2020, liao_human-centered_2022}. Drawing on work design literature~\cite{parker_smart_2024}, extending on Self-Determination Theory~\cite{ballou_self-determination_2022, ryan_self-determination_2000,ryan_self-determination_2018}, and in alignment with effective oversight requirements~\cite{sterz_quest_2024}, \citet{faas_design_2026} proposed design considerations for ``good'' oversight that span across five characteristics: \textbf{Stimulating}, \textbf{Mastery}, \textbf{Autonomous}, \textbf{Relational}, \textbf{Tolerable} (\autoref{fig:smartOversight}):

\begin{figure}[t]
    \centering
    \includegraphics[width=\linewidth]{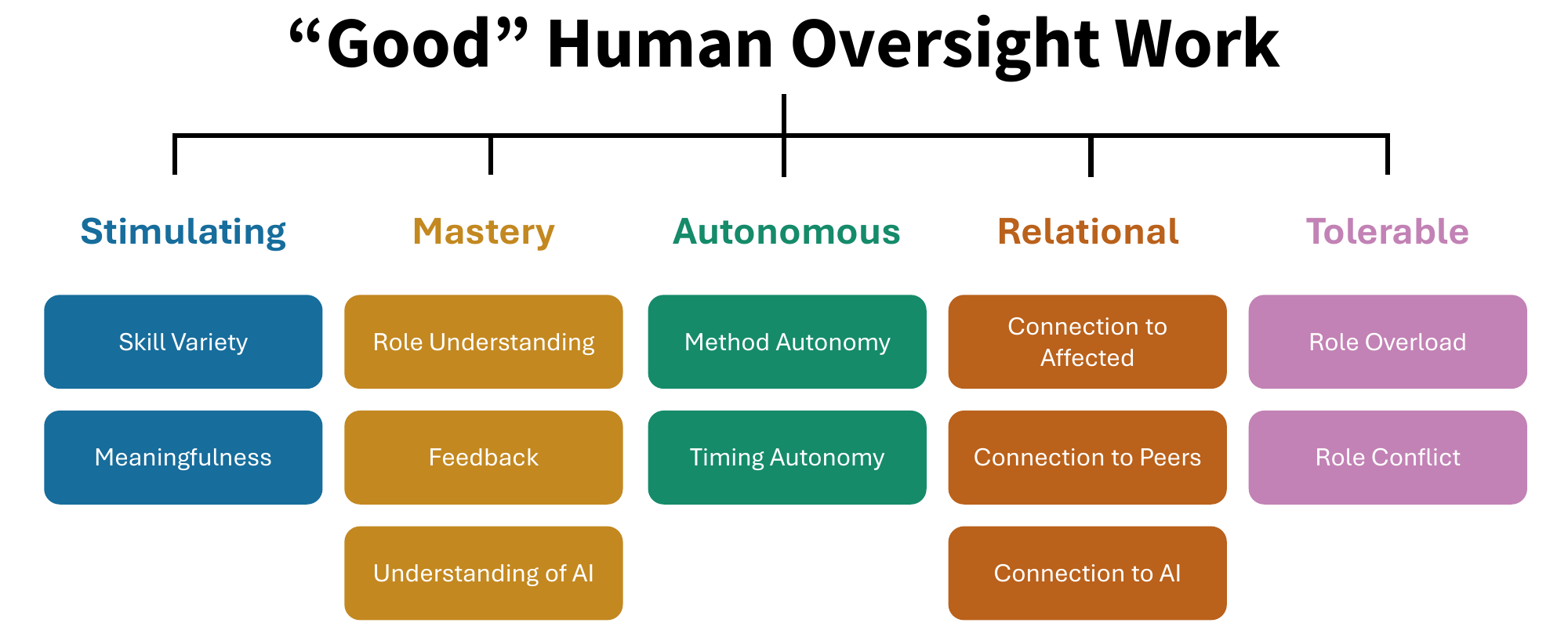}
    \caption{Design considerations that \citet{faas_design_2026} identified to lead to ``good'' human oversight work.}
    \label{fig:smartOversight}
    \Description{
    Overview of the design considerations of \citet{faas_design_2026}: Stimulating, which considers skill variety and meaningfulness. Mastery, which considers role understanding, feedback, and understanding of AI. Autonomous, which consists of method and timing autonomy. Relational, which consists of connection to affected, peers, and AI. Tolerable, which considers role overload and role conflict.
    }
\end{figure}

\begin{itemize}
    \item \textbf{Stimulating:} the design consideration require ``good'' oversight work to involve a \textit{variety of skills} (achieved e.g., through cognitive forcing~\cite{bucinca_trust_2021, miller_explainable_2023, reicherts_ai_2025}) and \textit{provide the overseer with meaning} as discussed in meaningful work literature~\cite{smids_robots_2020, bankins_ethical_2023, ranjit_are_2026}.
    \item \textbf{Mastery:} design considerations concern the overseer's \textit{understanding of the AI} and its behaviour, the \textit{feedback} the overseer receives about their performance, and their own \textit{understanding of their role} within the interaction~\cite{faas_design_2026, shome_why_2026}. While AI understanding is at the core of XAI research~\cite{longo_explainable_2024, ali_explainable_2023}, both feedback and role understanding can be supported by suitable interaction patterns~\cite{miller_explainable_2023, gomez_human-ai_2025, ma_towards_2025} and tackling the problem of de-skilling~\cite{ehsan_future_2026, passalacqua_practice_2024, caosun_augmentation_2026, shen_how_2026}.
    \item \textbf{Autonomous:} ``good'' oversight provides control over \textit{how} (\textit{method autonomy}) and \textit{when} to do tasks (\textit{timing autonomy})~\cite{faas_design_2026}. For decades, autonomy has been promoted in design guidelines~\cite{friedman_value-sensitive_1996, amershi_guidelines_2019} and supporting it through input modalities~\citep{Rogers_2010}, meaningful decision choices~\citep{Judge_2010}, or control over final outputs~\citep{young-ng_balancing_2025} is considered beneficial, while a lack of autonomy is associated with harms~\cite{bennett_how_2023}.
    \item \textbf{Relational} oversight should provide a \textit{meaningful connection to the people affected} by the outcomes~\cite{faas_design_2026, smids_robots_2020} and prevent workplace loneliness~\cite{bryan_loneliness_2023} by \textit{supporting human relationships}~\cite{faas_design_2026, boyd_artificial_2026, bieber_potential_2025, xia_impact_2025}. Further, recent research warns about various risks~\cite{marchegiani_anthropomorphism_2025, alberts_computers_2024, kirk_why_2025} associated with the \textit{desire to meet relational needs in human-AI interactions}~\cite{faas_design_2026, bieber_potential_2025, machia_social_2024}, reflected in AI anthropomorphization~\cite{akbulut_evaluating_2026, seymour_systematic_2023, xu_deep_2022} and human-AI relationships~\cite{moussawi_user_2018, brandtzaeg_my_2022, skjuve_longitudinal_2022}. 
    \item \textbf{Tolerable:} oversight considers the \textit{task load}, which was expected to be reduced with AI automation in place~\cite{amershi_guidelines_2019, green_principles_2019, endsley_ironies_2023, sellen_rise_2024}, but found to be condensed and intensified~\cite{dhanorkar_human_2026, chen_code_2026, tankelevitch_metacognitive_2024}, and \textit{role conflicts} with the AI, which should be avoided by clearly defined roles and boundaries~\cite{faas_design_2026, gomez_human-ai_2025, baum_constitutive_2026}.
\end{itemize}

The relevance of empirical evidence about these characteristics extends beyond interaction design and overseers' well-being: it has been identified as one of the dimensions along which compliance with human oversight requirements will have to be tested~\cite{langer_complexities_2026}. Most prior research, however, remains theoretical, offering design requirements without empirically testing them. Empirical work, meanwhile, often addresses human-AI interaction broadly rather than oversight-specific challenges, typically in short-term lab studies. \citet{ehsan_future_2026}, who studied real-world AI deployment in radiation oncology in a year-long study, and \citet{dhanorkar_human_2026}, who studied the lived reality of developers overseeing software agents in expert interviews, are notable exceptions, offering findings from real-world tasks performed over extended periods of time. But, like most studies, they focus on a subset of SMART characteristics and mainly discuss deteriorating effects on mastery and autonomy, while missing the discussion of stimulation, relatedness, and the tolerability of the oversight tasks.
Our study investigates all SMART characteristics longitudinally in oversight work performed in participants' everyday jobs.

\section{Method}
This paper explores the work characteristics of oversight work in a three-month longitudinal field study and how the design of oversight interfaces can support them.
Specifically, we took an action research approach and integrated in-the-wild observations with requirements elicitation workshops and work design theory. We accompanied tutors overseeing an AI-supported learning assistant (ALA) used by students of a university statistics lecture over a full semester. Through reflexive thematic analysis of two iterations of structured workshops enriched with insights from twelve weekly meetings and four repeated questionnaires, we identified key challenges in the oversight work. We then designed UI modifications grounded in design considerations for ``good'' oversight work~\cite{faas_design_2026}. Tutors used the modified interface during the final weeks, enabling us to explore how the user interface influenced the oversight work characteristics and tutors' task perception.

\subsection{Oversight Task}
\begin{figure}[t]
    \centering
    \includegraphics[width=1\linewidth]{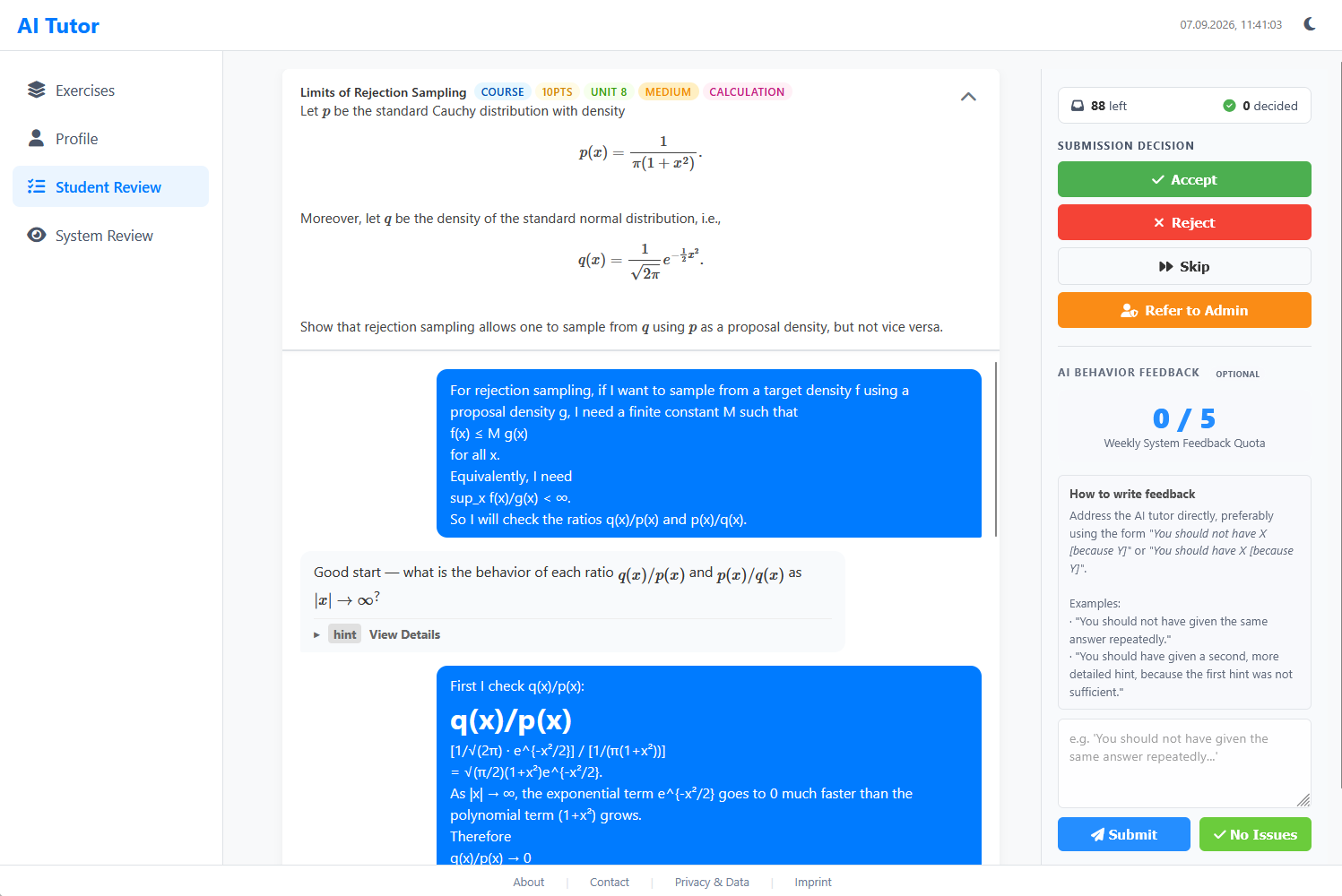}
    \caption{Screenshot of the oversight interface, including a conversation log and instructions to enter normative feedback on the AI's behaviour.}
    \label{fig:oversightInterface}
    \Description{
    Screenshot of the oversight interface, where the conversation between the student and the AI is displayed in the centre. On the right side, the tutors can enter their normative feedback on the AI behaviour. There are also instructions and some examples on how to provide normative feedback.
    }
\end{figure}

The Statistics lecture at the authors' University introduced an AI-supported learning assistant (ALA) implemented within a chat interface. Students used ALA weekly to discuss and solve dedicated exercises. The job of tutors was --- besides holding weekly tutorial sessions and grading exercises --- to oversee how ALA communicated with the 159 students by reviewing anonymised conversation logs, identifying undesired behaviour of ALA, and providing normative feedback on at least five conversations per week (see \autoref{fig:oversightInterface}). Feedback of this kind can shape a system's behaviour without retraining the underlying model~\cite[e.g.,][]{jahn_grace_2026}.

While, in principle, this form of human feedback could directly be fed back into the model, in this case, the lecturer first wanted to establish the quality of tutors' normative feedback and how it might change the assistant's behaviour before exposing students to feedback-driven changes. Thus, tutors' collective normative feedback was only used once to update the model as part of the \emph{action} in our action research cycle (see below, ~\autoref{sec:oversightInterface}).

\subsection{Study Design and Data Collection}
\begin{figure}[t]
    \centering
    \includegraphics[width=0.9\linewidth]{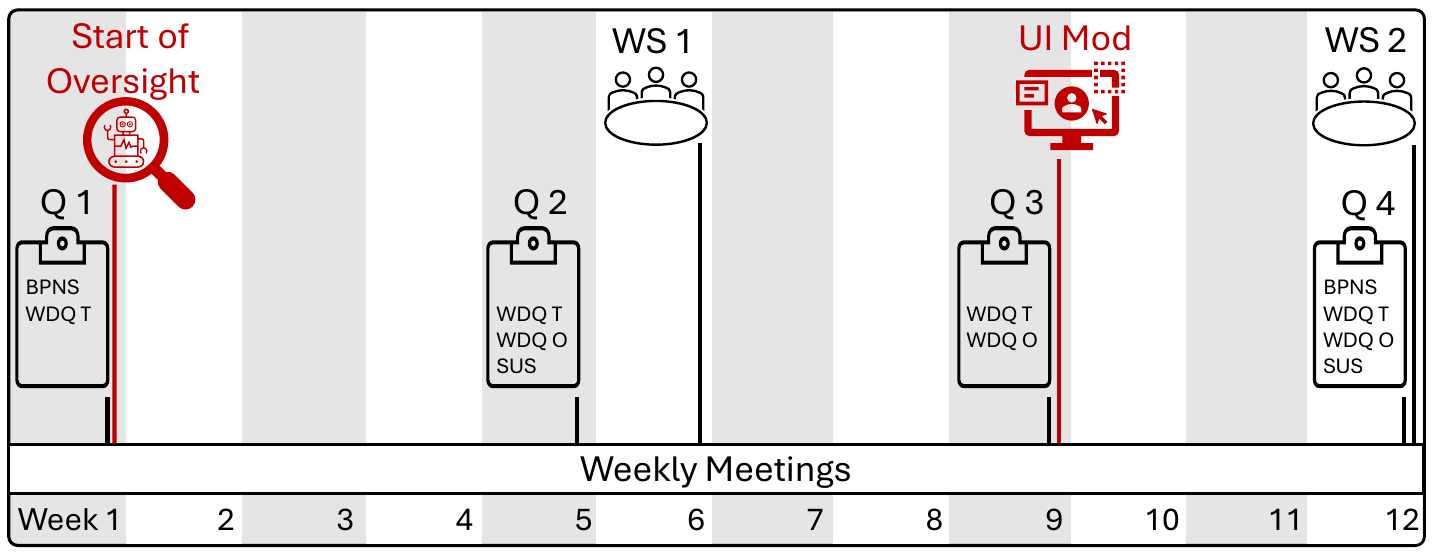}
    \caption{Overview of the study design. We performed four repeated questionnaires (Q1-4), two workshop iterations (WS1-2), and one UI modification.}
    \label{fig:studyDesign}
    \Description{
    The study spanned a 12-week period with weekly meetings. At the end of Week 1, participants filled out the first questionnaire, which assessed BPNS and WDQ T. After answering Q1, participants started to perform the oversight task. In Week 5, participants filled out the second questionnaire, which assessed WDQ T, WDQ O, and SUS. In Week 6, the first workshop iteration was performed. In Week 9, participants filled out the third questionnaire, which assessed WDQ T and WDQ O. After filling out the third questionnaire, the UI modifications were introduced. In Week 12, participants filled out the last questionnaire, which assessed BPNS, WDQ T, WDQ O, and SUS. After filling out the last questionnaire, the second workshop iteration was performed.
    }
\end{figure}

We accompanied tutors of a computer-science statistics lecture at the authors' institution over a three-month lecture period (see \autoref{fig:studyDesign}). Following an action research approach~\cite{hayes_relationship_2011}, we aim to understand the problems of human oversight in the tutor's everyday work. We target contextualised UI modifications that are transferable to similar problems despite not being generalisable. This paper reports one cycle of the action research spiral~\cite{lewin_action_1946}: an exploratory phase to understand tutors' main challenges and \textit{plan} supportive UI modifications, we then \textit{act} by deploying these modifications, and \textit{fact-find} through their effects. Our discussion (\autoref{sec:discussion}) outlines the next planning phase for future lecture iterations.

We accompanied tutors throughout the lecture period in two ways. 
First, one author attended the tutors' weekly one-hour meetings with the teaching assistants of the lecture, passively observing discussions about the tutor's work more broadly and how the oversight task affected it,  noting relevant comments without steering or participating in the discussions. These meetings were not recorded to keep them natural and uninterrupted. 
Second, we ran two one-hour workshops (after 6 and 12 weeks), in which participants first reflected individually on their experiences doing the oversight task, then built an affinity diagram of their experiences in groups of four. We video-recorded the affinity diagram and audio-recorded discussions. Workshops were conducted in English, and some quotes were shortened for clarity.
Importantly, to enable open and unbiased discussions with the tutors, the authors were not part of the team that organised and held the lecture.

We further collected repeated questionnaires on participants' need fulfilment (BPNS~\cite{gagne_role_2003, ryan_self-determination_2000}), motivational work characteristics~(WDQ following~\citet{parker_smart_2024} and adapted to tutoring job (WDQ~T) or oversight task (WDQ~O); with adopted item from~\cite{morgeson_work_2006, haynes_measures_1999, netemeyer_development_1996, mullarkey_employee_1995}), and the usability of the interface (SUS~\cite{brooke_sus_1996})  (see \autoref{fig:studyDesign} for overview and \autoref{app:QuestionnaireItems} for questionnaire items). 

Participants were introduced to the oversight task after they filled out Q1, and the second workshop was performed after Q4 was assessed.
Given our focus on the oversight task's work characteristics, we report only answers measuring \textit{Stimulation}, \textit{Mastery}, \textit{Autonomy}, \textit{Relatedness}, and \textit{Tolerability} (for need fulfilment and usability results see \autoref{app:needs} and \autoref{app:sus}). The study was approved by the authors' institution's ethical review board. Students were informed at the start of the lecture period that their conversations with ALA would be reviewed by tutors for quality-assurance purposes, and could opt out by declining consent to use their data for study purposes in the consent form shown on the first login. The users were also informed that they could revoke consent at any time. Conversations appeared in the oversight interface without student names or other identifiers. No student conversation data is included in the published data, which comprises questionnaire responses, meeting notes, and workshop transcriptions and is available on \href{https://zenodo.org/records/22247899?preview=1&token=eyJhbGciOiJIUzUxMiJ9.eyJpZCI6Ijk2MzNkYzczLTdmYWYtNDAzOC1hZTVhLWNlMzRiMTFmMjlmZCIsImRhdGEiOnt9LCJyYW5kb20iOiIwNWQ1YWRmNGVjMzc2OTFhY2MxNjQyYTgzMmY5ZThiYiJ9.EZAwkjC5IKaor76MqI9z7OQZe53C-sfWiciuAC2myD9IZAPbCxpZAjFWvAa-sIRzUM39cDu-1HXO_8v7C34ZJQ}{zenodo}; affinity diagrams are in \autoref{app:affinityDiagrams}. 

\subsection{Participants}
We conducted this study with tutors (\textit{N}=8, 7 male, 1 female, age 20–43, \textit{M}=24.87, \textit{SD}=7.56) of a computer science statistics lecture at the authors' institution. All tutors were advanced students in computer-science-related fields, half had prior tutoring experience, and all had the expertise required to tutor the statistics lecture. On a five-point Likert scale (1: do not agree at all, 5: agree completely), participants reported experience with AI tools (\textit{M}=4, \textit{SD}=0.71, [1–5]) but were indecisive about their understanding of AI (\textit{M}=3.5, \textit{SD}=0.71, [1–5]) and LLMs (\textit{M}=3.25, \textit{SD}=0.97, [1–5]).

\subsection{Thematic Analysis}
While we collected quantitative and qualitative data through the described questionnaires, workshops, and observations, our analysis focuses on qualitative insights. Questionnaire responses were only treated as complementary evidence, given that the sample size of 8 participants is insufficient for statistical hypothesis testing. Our main approach is a reflexive thematic analysis~\cite{braun_using_2006}, comprising five steps:

\begin{enumerate}
\item \textbf{Familiarisation:} we transcribed workshop recordings, digitised affinity diagrams, and immersed ourselves in data.
\item \textbf{Initial codes:} we generated initial codes from affinity diagrams, transcripts, and meeting notes, via data-driven semantic analysis closely representing participants' own words.
\item \textbf{Initial themes:} we integrated the high-level characteristics~\cite{faas_design_2026} of ``good'' oversight by grouping codes and forming initial data-driven themes within each.
\item \textbf{Reviewing themes:} we refined codes within themes, complemented semantic analysis with latent analysis where appropriate to capture underlying ideas and assumptions.
\item \textbf{Combining Analysis and naming themes:} after performing steps 1-4 for each workshop iteration individually, we named and refined each theme by combining the findings and integrating them with design considerations for ``good'' oversight~\cite{faas_design_2026} where appropriate.
\end{enumerate}

These steps reflect a hybrid approach: we used SMART characteristics as deductive frame and coding inductively within it. We aimed for detailed, nuanced insights and a rich account of which characteristics were most important to participants. Coding and initial themes were created by two co-authors (one of whom ran the workshops) and refined by all authors to increase analytical rigour and interpretative diversity.

\begin{figure}[p]
    \centering
    \begin{subfigure}{.47\linewidth}
        \centering
        \includegraphics[width=\linewidth]{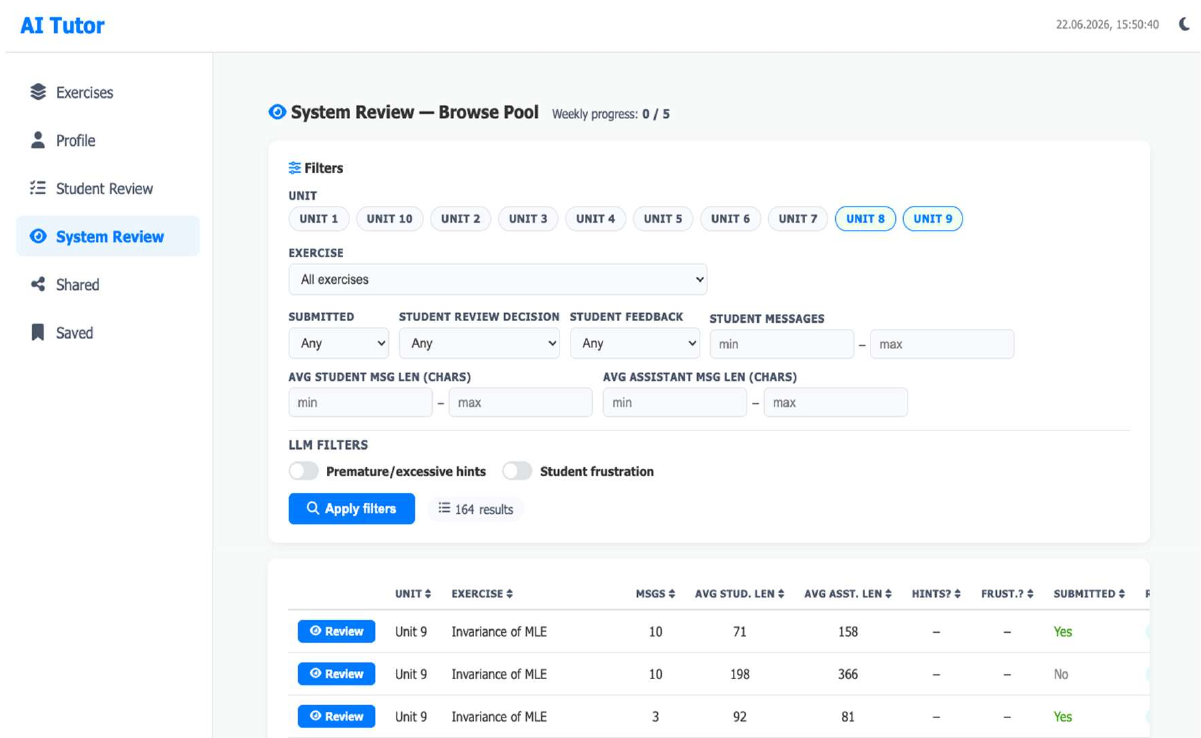}
        \caption{\filter}
        \label{fig:filter}
        \Description{
        A filter system was added, allowing the tutors to select interactions based on different criteria.
        }
    \end{subfigure}
    \begin{subfigure}{\linewidth}
        \centering
        \includegraphics[width=\linewidth]{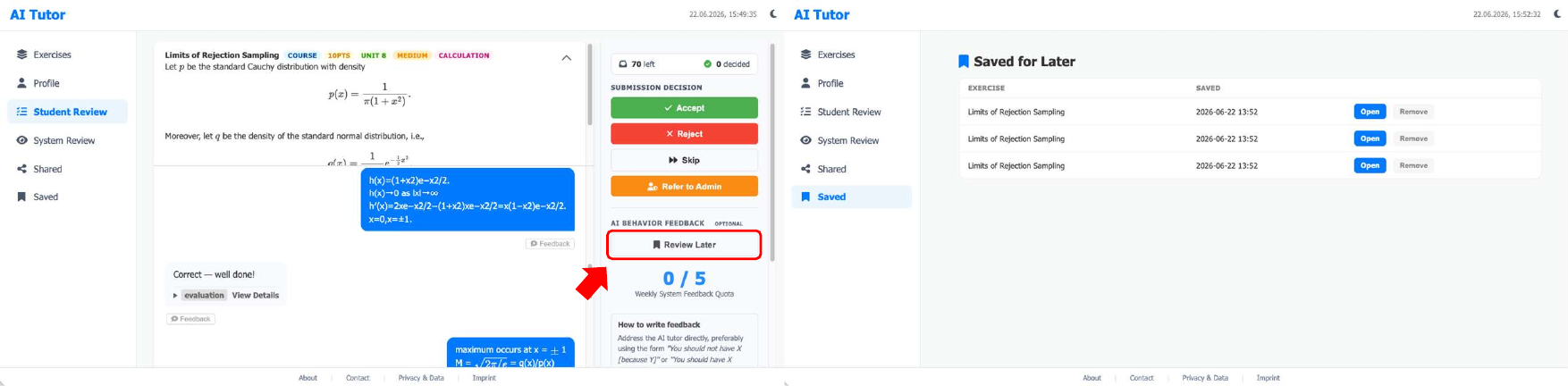}
        \caption{\save}
        \label{fig:save}
        \Description{
        A button was added labelled ``Review Later'' and a ``Saved''tab where the saved interactions can be found again.
        }
    \end{subfigure}
    \begin{subfigure}{\linewidth}
        \centering
        \includegraphics[width=\linewidth]{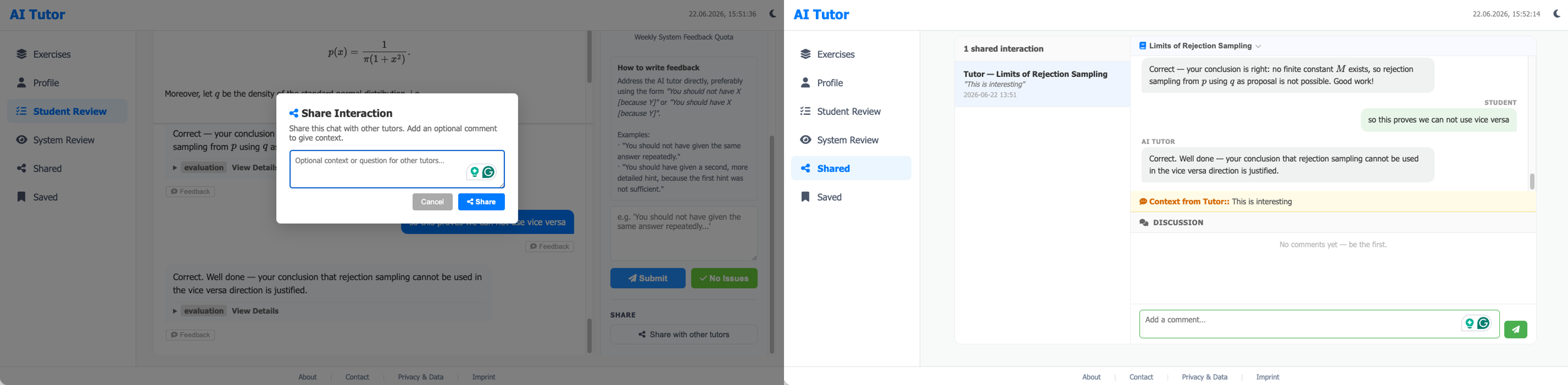}
        \caption{\share}
        \label{fig:share}
        \Description{
        A button was added labelled Share with other tutors'', which allowed sharing an interaction alongside a message. There was also a ``Shared'' tab added to view interactions that were shared among tutors and the discussion they have about the interaction.
        }
    \begin{subfigure}{.49\linewidth}
        \centering
        \includegraphics[width=\linewidth]{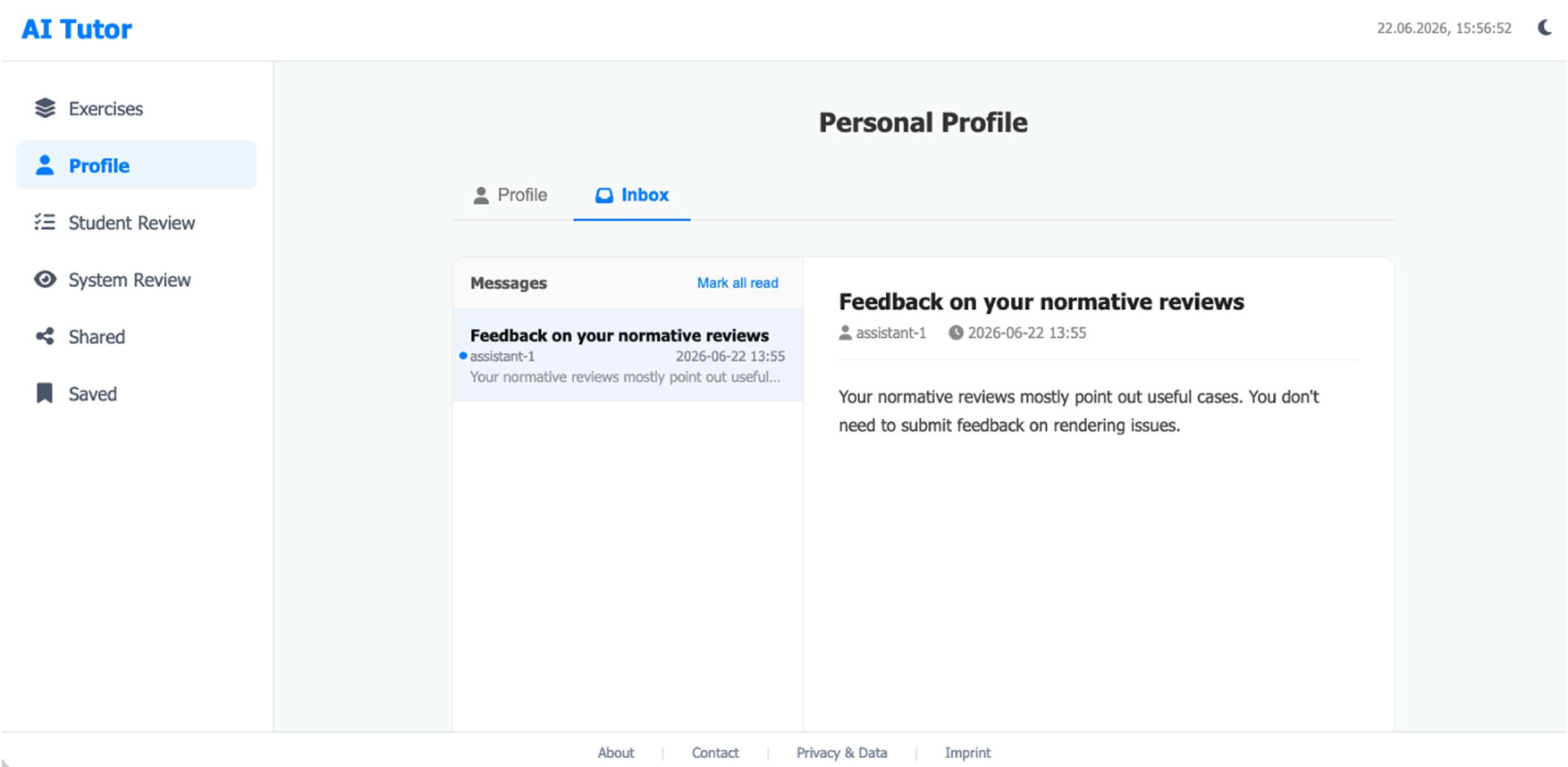}
        \caption{\tutorFeedback}
        \label{fig:feedback}
        \Description{
        An inbox was added to the tutor's profile tab, where they could receive feedback on their normative reviews from the teaching assistants.
        }
    \end{subfigure}
    \begin{subfigure}{.49\linewidth}
        \centering
        \includegraphics[width=\linewidth]{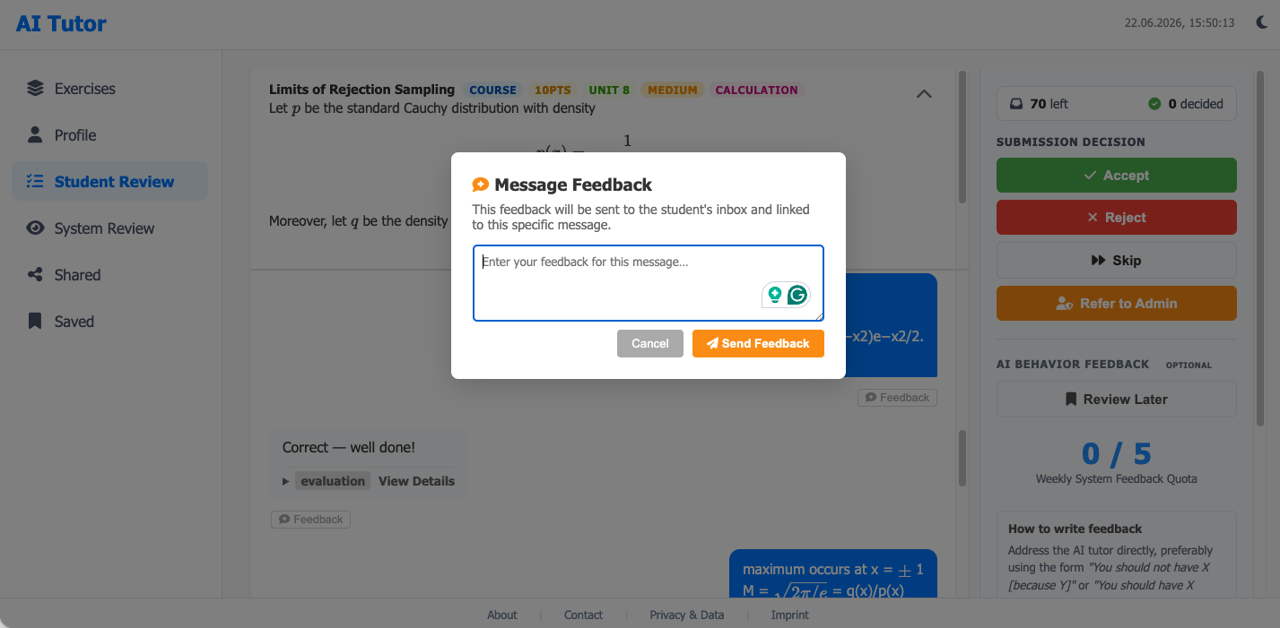}
        \caption{\studentFeedback}
        \label{fig:message}
        \Description{
        For every message of the AI tutor, there was a button added labelled ``Feedback'' which allowed sending a message directly to the student who performed the interaction.
        }
    \end{subfigure}
    \end{subfigure}
    \caption{Changes made to the user interface that aim to improve work characteristics of human oversight.}
    \label{fig:UIchanges}
    \Description{
    Screenshots of the UI modifications added to the oversight interface.    
    }
\end{figure}

\subsection{Interface Modifications}
\label{sec:oversightInterface}
After the first workshop, we analysed how participants experienced the oversight task so far. Informed by design considerations for ``good'' oversight work~\cite{faas_design_2026} and our initial findings, we designed UI modifications to compensate for potential deteriorating effects and to further investigate them\footnote{Claude Code with Anthropic's Sonnet and Opus Models was used to support the implementation of these features. All outputs were reviewed and verified by the authors.}. Since the workshop revealed how interlinked the characteristics are, some modifications targeted a single characteristic while others addressed multiple (see \autoref{tab:featureOverview}).

A \textbf{\filter} lets tutors build their own review pool, based on criteria such as chat length or average message length (\autoref{fig:filter}), aimed to support autonomy, stimulation, and tolerability. Similarly, \textbf{\save} allows marking conversations to be stored for later review, aiming to support tolerability and autonomy (\autoref{fig:save}).
Through the  \textbf{\tutorFeedback} (\autoref{fig:feedback}), tutors received comments on the normative feedback they give, which aimed to foster mastery. With \textbf{\share}, tutors can share an interaction with a comment during review, which appears on a discussion page visible to all tutors (\autoref{fig:share}), aiming to support relatedness with peers. Similarly, supporting relatedness to students, \textbf{\studentFeedback} allows directly resolving misconceptions, AI mistakes, or other concerns by sending a message to the student's inbox (\autoref{fig:message}). When the new interface was introduced in week 9, we additionally induced an \textbf{\AIbehavior} by distilling recurring themes from their previous normative feedback into a revision of the system prompt, aiming to increase stimulation and mastery by enabling tutors to experience the impact of their normative feedback. Based on the normative feedback that tutors have provided up to this point, these prompt changes mainly instructed ALA to keep context explanations short and to provide less-telling hints.

\begin{table}[h]
    \centering
    \caption{Features added to the oversight interface after the first two months, and which SMART characteristic they aimed to support.}
    \label{tab:featureOverview}
    \begin{tabular}{lccccc}
        \toprule
        UI modification & Stimulating & Mastery & Autonomous & Relational & Tolerable \\ 
        \midrule
        \filter & $\bigcirc$ & & $\bigcirc$ & & $\bigcirc$\\
        \share & & & & $\bigcirc$ & \\
        \tutorFeedback & & $\bigcirc$ & & &  \\
        \save &  & & $\bigcirc$ & & $\bigcirc$\\
        \studentFeedback & & & & $\bigcirc$ &  \\
        \midrule
        \AIbehavior & $\bigcirc$ & $\bigcirc$ & & & \\
        \bottomrule
    \end{tabular}
\end{table} 
\section{Results}
\label{sec:oversightTask}
\begin{figure}[t]
    \centering
    \includegraphics[width=\linewidth]{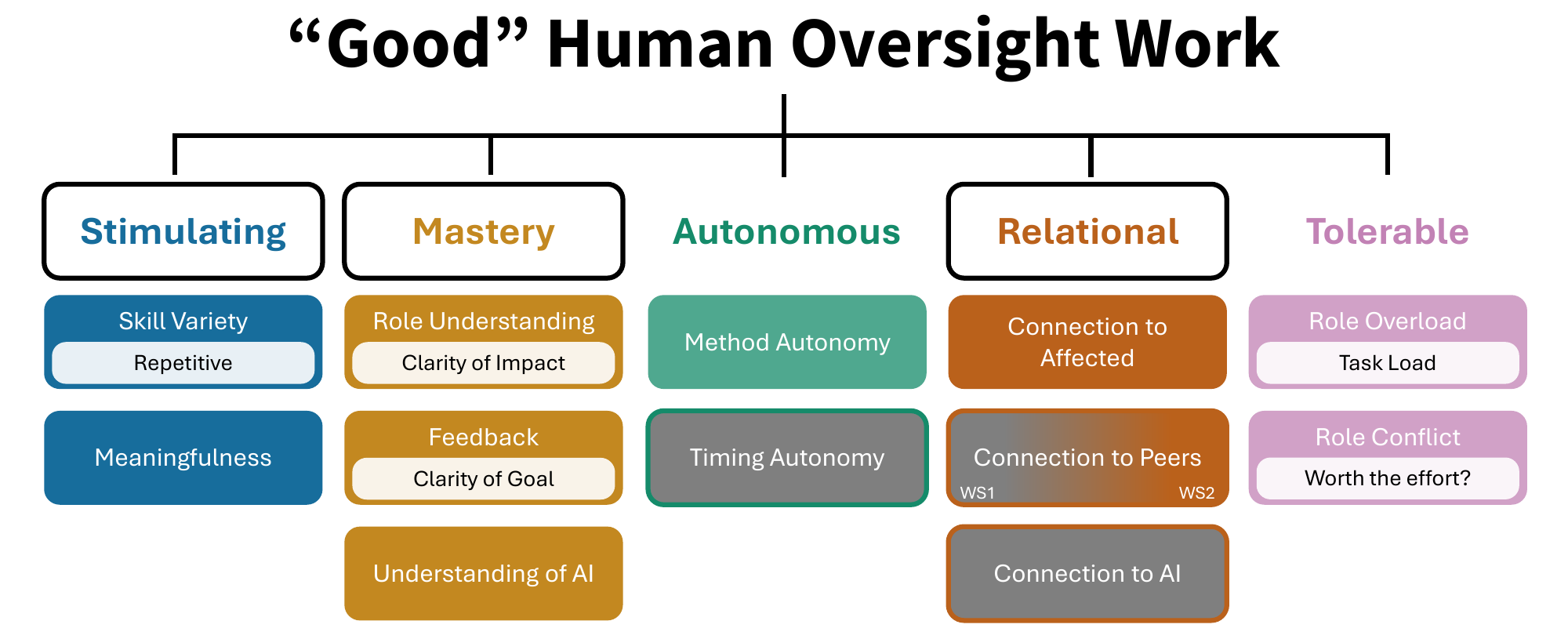}
    \caption{Overview of the thematic analysis that emerged from the workshops. Most prominent characteristics are highlighted, factors that did not emerge in our analysis are greyed out, and differing theme names are added for factors that emerged differently in our analysis.}
    \label{fig:thematicAnalysis}
    \Description{
    Overview of the findings of our thematic analysis adopted from the previous overview of design considerations of \citet{faas_design_2026}: Stimulating (highlighted), which considers skill variety, which was expressed as repetitiveness in our analysis, and meaningfulness. Mastery (highlighted), which considers role understanding, which was expressed as clarity of impact in our analysis, feedback, which was expressed as clarity of goal in our analysis, and understanding of AI. Autonomous, which consists of method autonomy and timing autonomy. In our analysis, only method autonomy emerged. Relational (highlighted), which consists of connection to affected, peers, which only emerged in Workshop 2, and AI, which did not emerge in our analysis. Tolerable, which considers role overload, which emerged as task load in our analysis, and role conflict, which emerged as ``worth the effort?'' in our analysis.
    }
\end{figure}

This section discusses the themes that emerged from the data within each of the SMART work categories, enriched with the responses from the repeated questionnaires (\autoref{app:wdq}). \autoref{fig:thematicAnalysis} visualizes our main themes within the SMART Oversight framework presented in~\autoref{sec:theoreticalBackground}. Overall, participants perceived the task as \textit{autonomous} and \textit{tolerable}, with these characteristics receiving less attention during discussions and receiving consistently high ratings in questionnaires. In contrast, \textbf{\textit{Stimulation}}, \textbf{\textit{Mastery}}, and \textbf{\textit{Relatedness}} were most affected. Below, we discuss how these characteristics emerged and evolved over time, the specific themes that came up within each characteristic, and whether UI modifications supported them.

\begin{figure}[ht]
    \centering
    \includegraphics[width=0.8\linewidth]{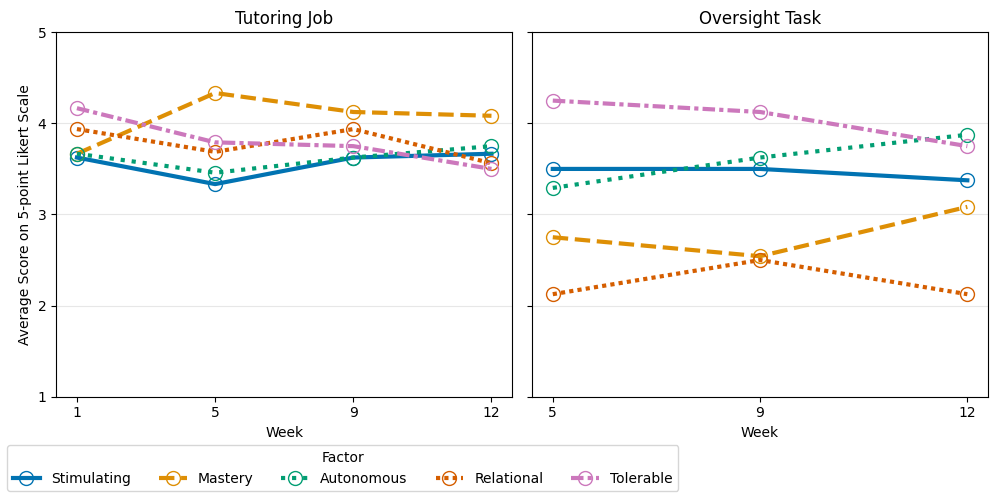}
    \caption{Work Characteristics of the Tutoring Job and Oversight Task on a Five-Point Likert Scale.}
    \label{fig:questionnaireSMART}
    \Description{
    Plots of the average score of work characteristics (SMART) of the tutoring job and oversight task on a 5-point Likert scale. For both the tutoring job and the oversight task, stimulation was rated at approximately 3.5 throughout the lecture period. Mastery of the tutoring job increased from 3.6 in Q1 to 4.1-4.3 in the other questionnaires. For the oversight task, mastery evolved from 2.8 to 2.6 to 3.1. Autonomy remained approximately constant at 3.6 for the tutoring job and increased continuously from 3.3 to 3.9 for the oversight task. Relatedness of the tutoring job went from 3.9 down to 3.7, back up to 3.9, and further down again to 3.6. For the oversight task, it increased from 2.1 to 2.5 and decreased back to 2.1. The tutoring job started at a tolerability of 4.2, then dropped to 3.8 in Q2+Q3, and further to 3.5 in Q4. For the oversight task, tolerability decreased from 4.3 to 4.1, then to 3.8.
    }
\end{figure}

\subsection{Stimulating} 
\label{subsec:Stimulation}
The task's \textbf{repetitive} nature was fundamental across both workshops; while likely impossible to eliminate, our UI modifications increased variety in monitoring and intervention. Further, participants judged the oversight task's \textbf{meaningfulness} either by improving the AI for students' benefit or by learning skills applicable elsewhere in their work. Where Workshop 1 revealed the task as mostly meaningless with only occasional meaningful moments, the UI modifications helped surface this meaning and sparked further ideas for weaving it more deliberately into oversight interactions.
This connects to the design considerations for ``good'' oversight by \citet{faas_design_2026} that advocate stimulating oversight, by requiring \textit{skill variety} and increasing \textit{meaningfulness}. Since our oversight task lacked these requirements, the task was perceived as repetitive, meaningless, and little stimulating. 

\subsubsection{Repetitiveness}
The task's \textbf{repetitive} nature (P1,3,4,5,6,7, WS1; P1-8, WS2) was consistently perceived as ``tiring'' (P6, WS1), ``boring'' (P3, WS1), ``tedious'' (P7, WS1), and ``like a chore'' (P3, WS2), mostly linked to the content of the overseen chats. However, after experiencing the \AIbehavior\ in \textbf{Workshop 2}, participants expected that a greater impact of their interventions on AI behaviour would make this repetitiveness more tolerable (\autoref{subsec:Tolerable}). 
In Workshop 2, participants valued the control \filter\ gave them over which chats to oversee, allowing focus on the most stimulating ones (P1-7). Still, oversight was perceived as repetitive, and participants wished for further improvements to increase variety: P5 and P6 suggested grouping chats to provide normative feedback in a bulk format and avoid repeating the same feedback.
Notably, the UI modifications increased variance in questionnaire answers about perceived stimulation, suggesting that the UI design could compensate for the deteriorating effect on stimulation for some tutors.

\subsubsection{Meaningfulness}
In \textbf{Workshop 1}, participants derived stimulation from the \textbf{meaningfulness} of cases where they expect their oversight to \textit{have an impact} (P1,2,6) or let them \textit{learn} (P5-7). P2 described providing normative feedback on misbehaviour as ``empowering'' and ``a dopamine hit''; P1 expected the task would ``feel better for us, in terms of productivity'' when their actions visibly changed the AI. P7 valued how it ``helps me reflect on my own help for students'', while P6 and P5 enjoyed learning from the prompts as a unique aspect of the task. While, in the first weekly meetings, participants gave the impression to see meaning in improving the system, P3 felt the task ``didn't feel like it was the job of a tutor'' and was ``time-consuming for the wrong reasons'', since helping students---not overseeing AI---constituted the tutoring job's meaning; P6 similarly found little value beyond personal learning (``I don't know how this particular thing is gonna help me in any way''). Accordingly, the oversight task was brought up less and less over the course of the lecture period. Interestingly, the questionnaire showed no clear stimulation difference between tutoring and oversight (\autoref{fig:questionnaireSMART}).

After 12 weeks and after experiencing the \AIbehavior, participants were clearer on the meaningfulness of their oversight task in \textbf{Workshop 2}. They framed their oversight's purpose as improving the AI (P1,5,7,8), but ultimately valued this for its downstream \textit{impact on future students} (P1,6,7,8; ``the main goal in the long term is to improve the ALA for the upcoming students,'' P8), alongside continued value in \textit{learning} (P4,6,8). UI modifications that reduced repetitiveness also helped participants focus on meaningful interactions, reinforcing that oversight felt meaningful only when it improved the AI for students' benefit or built skills applicable elsewhere in their tutoring role.

\subsection{Mastery}
\label{subsec:Mastery}
Before the UI modification, participants felt little mastery in the oversight task, especially relative to their tutoring job (see also \autoref{fig:questionnaireSMART}). This stemmed from uncertainty about three aspects: the \textbf{clarity of impact} of their interventions, the \textbf{clarity of their goal}, and their \textbf{understanding of the AI}, which recurred across both workshops. In Workshop 1, participants found their goal ambiguous and subjective, their impact unclear, and wished for better AI understanding. 
Counter to the usual expectation that mastery grows with practice, only one participant's questionnaire responses showed this pattern from week 5 to week 9, with some even reporting lower mastery in the second questionnaire. In the questionnaires, this reversed after the UI modifications: mastery increased for all but one participant (unchanged) between workshops (\autoref{fig:wdq-o}). Discussions in Workshop 2 revealed how all three aspects can be improved through interaction design, but also point at the limits of interaction design, which we will discuss in more detail in \autoref{subsec:designImpl}. 

The concerns that emerged here are also discussed in the design considerations for ``good'' oversight~\cite{faas_design_2026}: the tutors lack an \textit{understanding of the AI}, an \textit{understanding of their oversight role},  which is largely influenced by the impact the tutors can have through their interventions, and miss \textit{feedback} about their goal and performance.

\subsubsection{Clarity of Impact}
In \textbf{Workshop 1}, the biggest concern was missing \textbf{clarity of impact} of the provided normative feedback (P1-7): participants ``don't know whether it helps'' (P5) or ``if ALA is affected'' (P7), which P2 attributed to the impact of their interventions not being immediately visible. P6 imagined that knowing how their normative feedback helped would itself be ``a good skill''. This remained a long-term concern in \textbf{Workshop 2} (P1-8). In one weekly meeting, a tutor mentioned they had noticed the \AIbehavior, but their interventions did not continuously translate into ``actual improvements in the responses of ALA'' (P6), and that impact remained ``not always so visible'' (P5).  To overcome this problem, some participants used \studentFeedback\ as way of more direct and targeted intervention that increased the mastery they experienced and ``helped pinpoint when we had some problems with the AI'' (P3). While P6 used \studentFeedback\ as a direct intervention targeted at the students, which gave them ``more authority to judge actually where is it going right or where is it going wrong'', P5 used the feature in an undesired way by providing normative feedback to the AI, addressing errors in specific messages rather than the whole chat. 
Clarity of impact was thus a persistent concern: tutors used \studentFeedback\ to make more targeted and somewhat clearer interventions, but the uncertainty about whether and how their normative feedback mattered remained.

\subsubsection{Clarity of Goal}
Already in the first weekly meeting, after the oversight task was introduced, participants expressed concerns about missing guidance and an unclear oversight goal. In \textbf{Workshop 1}, participants perceived their goal as subjective (P2,4,5), making decisions difficult: ``it's very difficult [...] to find the balance'' (P4), and normative feedback feeling ``like giving an opinion'' (P2). In \textbf{Workshop 2}, the goal was framed (P1-4) as a lack of shared clarity rather than mere subjectivity: P3 noted that ``tutors really don't know what exactly is the goal for the AI to behave like at the end'', calling it ´´ambiguous'', while P2 ``realized that in one chat I see this problem [...] and then in another chat I see it as the opposite''. P1 accordingly wished for more guidance ``so that we are all on the same page'' and utilised \save\ to get this guidance from discussions with the teaching assistants or other tutors. While P7 valued the feedback they received about their interventions in \tutorFeedback, they wished it was more frequent and real-time. P4 expected a clear goal would improve ``fairness''.
The goal's ambiguity thus evolved from an individual subjectivity in Workshop 1 to an explicit call for shared, common guidance in Workshop 2.

\subsubsection{Understanding of AI}
In both workshops, participants wanted a better understanding of the AI and their own performance, to feel greater impact and mastery (P1,6,7, WS1; P5,6, WS2). In \textbf{Workshop 1}, P7 wished for a way of ``seeing the progress of ALA'' and P6 found it valuable to ``learn how LLMs work''. In \textbf{Workshop 2}, P5 noted that \share\ ``gives more understanding of the system itself'', though P6 maintained that deeper AI understanding was still needed to effectively intervene.

\subsection{Autonomous}
\label{subsec:Autonomy} 
Across both workshops, autonomy was not a prominently discussed characteristic, with only some concerns raised about \textbf{method autonomy}. Combined with the high questionnaire ratings (\autoref{fig:questionnaireSMART}), autonomy's low salience likely reflects that the oversight task, as set up here, already provided sufficient control and freedom, leaving other needs as more pressing concerns. Still, Workshop 2 showed that our UI modifications improved method autonomy and sparked further ideas for extending it.

While \citet{faas_design_2026} describe the same need for \textit{method autonomy} in their design considerations, they also require \textit{timing autonomy} to decide the timing of one's responsibilities freely. Since this was mentioned only by P3, who perceived high timing autonomy, we assume that the oversight task provided participants with sufficient timing and method autonomy.

\subsubsection{Method Autonomy}
In \textbf{Workshop 1}, participants experienced high autonomy in choosing method (P1,5,6,8) of their oversight, using this freedom to decide ``what goal to pursue'' (P1) and develop their own guidelines. P1 linked the task's lack of stimulation to limited method autonomy, noting they ``don't have very much to play with''. One restrictive aspect was the binary decision space, which P6 suggested extending for more granular decisions; P5 described working around this by referring some decisions to teaching assistants.
In \textbf{Workshop 2}, participants valued the freedom supported by the new \filter\ (P3,7), but still wished for more method autonomy (P3,5) like choosing to rate answers with a simple thumbs-up/-down and only ``adding additional [normative] feedback in words'' where needed (P3). P3 also wanted the choice between actively ``find[ing] specific chats'' or simply browsing ``what chats [are] given'', while P5 wanted to group chats to better direct their focus, echoing the earlier desire for more granular, structured control over the decision space.

\subsection{Relational}
\label{subsec:Relational}
Although the questionnaire data on Relatedness showed the largest difference between the oversight and tutoring tasks compared to other characteristics (\autoref{fig:questionnaireSMART}), low relatedness was not a core concern in Workshop 1 and only became prominent after experiencing \share\ in Workshop 2, which enabled a \textbf{sense of peer community} among tutors, and this experience further revealed a desire to \textbf{connect with students} through the oversight task---reinforcing that positively affecting students constitutes the task's core meaningfulness (\autoref{subsec:Stimulation}).
Questionnaire data complicates this picture, though: despite Workshop 2 discussions suggesting participants experienced the task as more relational, all but one participant rated it as equally or \textit{less} relational afterwards (\autoref{fig:wdq-o}). Participants may initially not have minded the oversight task's low relatedness, since the tutoring job overall was sufficiently relational. Once the \share\ revealed the potential for a more relational task, however, participants may have become more aware of, and more bothered by, how little relatedness it still provided---leading to equal or lower ratings.

Similarly, \citet{faas_design_2026} stress the importance of supporting \textit{relationships to peers} and \textit{relationships to affected individuals} through interface design. Interestingly, they further caution about the potential risks of desiring a \textit{relationship to the AI}, which did not emerge in our workshops. Therefore, we refrained from adding UI modifications that foster human-AI relationships.

\subsubsection{Connection to Affected}
In \textbf{Workshop 1}, participants stressed the importance of connecting with students in the tutoring job, either explicitly or implicitly (P6,7), or feeling disconnected from students in the oversight task specifically (P3,5,7,8). P6 described being ``engaged with the students'' as central to tutoring and a ``lifelong skill'' missing from oversight. P7 similarly valued guiding students as ``worthwhile'' and---unlike P6---also used the oversight task itself for self-reflection to improve their student support elsewhere. Still, P3, P5, P7, and P8 felt their oversight had ``no big impact'' (P5) on students, which left the tutors with the impression they ``don't actually care'' about the students (P3).
Relatedness was discussed less in Workshop 1 than the questionnaire data would suggest. We see two possible explanations: either the tutoring job's other tasks already sufficiently met participants' relatedness needs (plausible, given relatively high ratings), or participants implicitly desired more relatedness without articulating it directly---consistent with how meaning in tutoring was tied to the students, suggesting relational content may be implicit within stimulation and mastery (\autoref{subsec:Stimulation}, \autoref{subsec:Mastery}).
This changes in \textbf{Workshop 2}, after participants experienced the interface modifications. Connection to students resurfaced as an explicit extension of the \share\ feature (P3,6), closely tied to locating the task's meaning in student impact (\autoref{subsec:Stimulation}). P3 and P6 proposed extending this social feature to connect students and tutors directly. P3 envisioned students and tutors to discuss the shared goal for AI behaviour (that we found missing in \autoref{subsec:Mastery}) as ``a good opportunity to maybe connect like the students and tutors.''

\subsubsection{Connection to Peers}
Connection to peers was raised \textbf{only in Workshop 2}, after tutors experienced the UI modifications \share\ and \save. Participants valued the resulting support for peer connection (P1,5,6,7,8). P7 used \share\ for interesting chats, and P8 for ``any funny thing or any crazy thing''. P1 saw \save's main value in enabling later discussion: ``now I can save it for later, discuss it with either of the [teaching assistants] or with the other tutors''. 
Interestingly, participants disclosed that they already shared anonymised interactions with other tutors via private channels before \share\ was introduced. 
Overall, \share\ created ``a sense of community'' among the tutors (P5), and P6 proposed extending these social features to foster community with students too, linking to the desire for stronger connections to students discussed above.

\subsection{Tolerable}
\label{subsec:Tolerable}
Generally, participants found the task's mental and temporal demands manageable (also see \autoref{fig:questionnaireSMART}), but consistently linked its tolerability to its meaningfulness—about which they remained unsure throughout (\autoref{subsec:Stimulation}). We therefore believe tolerability in our use case depends more on meaningfulness than on \textbf{task load} itself.

This aligns with the design considerations for ``good'' oversight~\cite{faas_design_2026} that stress that \textit{role overload} should be prevented by presenting the task load in a manageable way to the users and \textit{role conflict} by clearly separating the tasks between the overseer and the AI. In our study, participants were getting frustrated since they perceived the AI to do all meaningful tasks, leaving them with worthless efforts.

This helps explain a puzzling contrast with the questionnaire data: although tolerability was even less prominent after the UI modifications were introduced and participants were more familiar with the task, questionnaire ratings somewhat decreased over time (\autoref{fig:questionnaireSMART}). Since participants closely linked tolerability to stimulation, we suspect the UI changes did not sufficiently increase stimulation to make the same task load feel more \textbf{worth the effort}. Additionally, there might be the confounding factor that task loads of both the tutoring job and lectures the tutors attend themselves as students intensify towards the end of the semester, which influences how tolerable they perceive the task load of the oversight task. Improving tolerability further will likely require making the task load feel worth it through meaningfulness, rather than reducing the load itself.

\subsubsection{Task Load}
Participants consistently discussed temporal (P1,3, WS1; P6,8, WS2) and mental demands (P1,4,6,8, WS1; P3,4, WS2) in \textbf{both workshops}. The task was described as ``time-consuming'' (P3, WS1; P6, WS2) and taking ``too much time'' (P1, WS1). The two were closely intertwined: P6 found the task ``quite tiring,'' and P4 reached a point of not wanting ``to read it anymore'', since participants had to read entire conversations (P1,6, WS1) perceived as ``too long'' (P4, WS1). Still, the task was, in theory, ``clear'' (P5, WS1) and ``structured in a way that is simple'' (P7, WS1), which made the load more tolerable (P3,5,7, WS1) even as some specific decisions remained difficult (P3,4, WS2).
Extending on the introduced UI modifications in \textbf{Workshop 2}, P3, P5, and P7 discussed how grouping chats could reduce the task's time cost and let them use their effort more efficiently. After 12 weeks of oversight mental demand of specific difficult decisions remained high, while P1 noted that \save\ made the task more tolerable.
Task load was thus a concern in both workshops, with clarity and structure expected as mitigators in Workshop 1 and further concrete measures in Workshop 2. In general, participants found the demands of our oversight task manageable in absolute terms.

\subsubsection{Worth the effort?}
From \textbf{Workshop 1} on, participants closely linked task load's tolerability to stimulation and relatedness. P3 put it directly: ``it is not the problem that it is time-consuming, but it is time-consuming for the wrong reasons''---implying the same load would feel more tolerable if the task were more stimulating and relational. This connects to participants' uncertainty about the task's meaningfulness (\autoref{subsec:Stimulation}), which shaped whether the load felt worth the effort.
This question wasn't resolved by the UI modifications in \textbf{Workshop 2}: P8 asked whether their ``reviews [would] be valued and considered by improving [the AI]?'', echoing the earlier concern that time spent on unclear purposes feels burdensome, and implied a role conflict: the AI should support them, yet ``ALA is not solving the problems for us''.
This tension was present in both workshops, with tolerability consistently framed as dependent on the tutor's stimulation, relatedness, and meaningfulness, rather than on the task load.

\section{Discussion}
\label{sec:discussion}
Our thematic analysis observed all characteristics of ``good'' oversight work identified by \citet{faas_design_2026}, but in line with their framework: while all characteristics matter, we investigated the main challenges of our oversight task. 
Due to the limited stimulation provided by their oversight task, participants perceived it as a repetitive chore that lacked meaningful impact. Closely linked, they perceived high uncertainty about their role in the oversight configuration, which was expressed through their uncertainty about the goal and the impact of their interventions and led to overall low mastery. 
Further, participants initially perceived their oversight task as little relational, and while this improved among peers, the lack of connection to students remained a persistent problem.
Participants' needs for autonomy were better met since their oversight task provided them with sufficient control and freedom. Similarly, they perceived the task as tolerable due to its manageable workload, but increasingly questioned whether this workload is worth the effort, given the lack of meaning they saw in the oversight task.

The deteriorating long-term work characteristics, as indicated by both workshop iterations and questionnaire responses, are consistent with prior research~\cite{faas_ai_2026, ehsan_future_2026}. 
Compensating for this remains critical, as they leave needs unfulfilled, which in turn harms job satisfaction, well-being, and performance in the long term~\cite{parker_smart_2024, Parker_2014_beyondmotivation, Humphrey_2007}. 
When work design fails to support these needs, workers may resort to unintended workarounds to meet their needs~\cite{Strich_2021, cheon_fulfillment_2025} as we observed with the unintended use of \save, \studentFeedback, and sharing anonymised interactions with other tutors via private channels before \share\ was introduced.

Overall, our findings align with prior work design, oversight, and human-AI interaction literature, contributing insights from a longitudinal field study into oversight's work characteristics in AI-supported education. Below, we discuss our observations in more detail, evaluating how our UI changes affected work characteristics and how ``good'' oversight design supports workers' psychological needs (\autoref{subsec:discussionUI}), extracting opportunities and limits of UI Design for supporting ``good'' oversight work (\autoref{subsec:designImpl}), and assessing how our observations transfer to other oversight tasks (\autoref{subsec:discussionOversight}).

\subsection{Success of Interface Modifications}
\label{subsec:discussionUI}
Our interface modifications (\autoref{sec:oversightInterface}) show opportunities for interaction design to support ``good'' work characteristics. In \autoref{tab:successUIchanges}, we revisit our hypotheses informed by our initial findings and design considerations for ``good'' oversight work~\cite{faas_design_2026} and discuss which characteristics the different UI modifications supported. Participants most valued the control gained from the \filter\ and the sense of community from \share. They also valued receiving the \tutorFeedback\ and observing the \AIbehavior\ based on their interventions, while they envisioned that more frequent \tutorFeedback\ and continuous \AIbehavior\ could further improve mastery and stimulation.
The \save\ and \studentFeedback\ features were discussed only briefly. Since the task was perceived as largely autonomous and tolerable, the need to save interactions for later was likely limited. We expect greater use in less autonomous or tolerable oversight tasks. Notably, one participant used \save\ to connect with others in person. This unintended use illustrates interface design's potential to support need fulfilment.
Similarly, there were limited discussions of how \studentFeedback\ supports relatedness, but more on how it was used in (un-)intended ways to intervene more directly or targeted. Two likely reasons for \studentFeedback\ being little relational are that interactions were anonymous and that students could not reply to the message. This aligns with participants' desire for more direct interaction with students through the oversight interface.

\begin{table}[h]
    \centering
    \caption{Overview of qualitative findings on the success of the design modifications, the characteristics they intend to support ($\bigcirc$), the characteristics we judge as successfully supported (\greenmark), which support we think requires higher frequency (\orangemark), and which characteristics we judge as less successfully than expected (?).}
    \label{tab:successUIchanges}
    \begin{tabular}{lccccc}
        \toprule
        UI modification & Stimulating & Mastery & Autonomy & Relational & Tolerable \\ 
        \midrule
        \filter & \greenCircMark & & \greenCircMark & & \greenCircMark\\
        \share & & & & \greenCircMark & \\
        \tutorFeedback & & \orangeCircMark & & & \\
        \save & & & \CircQuestionmark & \greenmark & \CircQuestionmark\\
        \studentFeedback & & \greenmark & & \CircQuestionmark & \\
        \midrule
        \AIbehavior & \orangeCircMark & \orangeCircMark & & & \\
        \bottomrule
    \end{tabular}
\end{table} 

\subsection{Opportunities and Limits of UI Design for Supporting ``Good'' Oversight Work}
\label{subsec:designImpl}
Importantly, our findings show that oversight work characteristics can be improved through focused interaction design. 
Through our action research approach, we identified tutors' main challenges as limited \textbf{\textit{stimulation}}, \textbf{\textit{mastery}}, and \textbf{\textit{relatedness}}. The next step in our action-research would be a consolidation of the introduced features, which is in line with participants' own preference for a smaller set of features used more often, such as extending \filter\ and \share, updating \tutorFeedback\ and \AIbehavior\ more frequently, and dropping the other features. 
The tutors themselves identified further opportunities for interaction design, which we discuss below, along with expected problems tied to this configuration of human oversight (\autoref{subsec:discussionOversight}). 

While repetitiveness seemed closely linked to the task's nature, tutors envisioned more design solutions: grouping chats by similarity, so normative feedback is needed per group rather than per chat, or a faster intervention mode for some conversations (e.g., thumbs up/down). Combining these, tutors could label each piece of normative feedback and reference it when similar misbehaviour recurs, still requiring monitoring of repetitive interactions but increasing the uniqueness of each intervention. To decrease the repetitiveness of monitoring, oversight could be performed on a higher level: interactions could be grouped and their behaviour summarised. This trades sampling density for variety, which depends on the trustworthiness and accuracy of the summarisation method.
We expect more frequent \tutorFeedback\ and \AIbehavior\ updates to substantially improve perceived mastery, though the indirect intervention remains tied to the oversight's asynchronous nature. In \autoref{subsec:discussionOversight}, we discuss how other oversight forms address this downside and the trade-offs involved.
Finally, participants envisioned extending ALA into a social platform for students and tutors alike. Extending \share\ could let tutors connect with students directly---either by discussing shared interactions with them or by overseeing interactions students share.

One design direction our participants pointed to deserves a caveat. Several valued the oversight task for what it taught them---about how the assistant answers (P5, P6, \autoref{subsec:Stimulation}). Designing for this is attractive: it addresses stimulation and it is the one source of meaning tutors reported that does not depend on closing the loop from performing an intervention to observing its impact. But it creates a dependency of the overseer's task-relevant judgments on the system they are meant to assess~\cite{baum_constitutive_2026}. Making oversight work more rewarding by making it more instructive may create a design tension with the capacity for (at least partial) independent judgment that makes oversight worth having. 

Our participants' \textit{mastery} concern is not, we think, a role-clarity deficit that better interaction design would resolve. They desired more fairness and reported an ambiguous goal, with insufficient guidance, leading to contradicting judgments. Thereby, they had located a \textit{normative gap}, not an informational one. Normative feedback, as they were asked to give, is evaluative: it says ALA should have behaved otherwise, judging against an unspecified standard. 
From the measurement side, the overseer's ability to detect errors cannot be characterised without a fixed standard for what counts as an error~\cite{langer_effective_2025}, which has to be chosen rather than discovered for normatively loaded judgments. 
From the other side, combining divergent normative judgments into system behaviour requires choosing an aggregation procedure, which is a normative decision rather than a neutral technical step~\cite{baum_aggregation_2025}. 
Our action research surfaced this problem in the field, with tutors producing mutually inconsistent normative feedback, noticing the inconsistency, and asking for a common standard so that the outcome is fair.

This bounds what interface design can do here. An interface can make the prevailing standard visible, show tutors where their judgments diverge from their peers', and support the discussion in which a standard is negotiated---our \share\ and \save\ features turned out to do a little of the third by accident. What it cannot do is supply the standard. Somebody---the teaching team, the deployer, the provider (in a regulated setting)---has to decide what the assistant ought to do, and own that decision. If this is left unassigned, interface design cannot fill this normative gap and delegates a normative choice to whichever overseer happens to be reviewing the conversation.

\subsection{Transferability to Other Configurations of Human Oversight}
\label{subsec:discussionOversight} 
Our action research approach aimed more at transferability to similar oversight configurations than at generalizable insights. We studied an asynchronous oversight task, and while some insights may extend to other forms of human oversight or human-AI interaction more broadly, caution is warranted, as transferability is linked to other tasks' similarity, and different work characteristics may emerge even in seemingly similar tasks. Since we expect an oversight person's work characteristics to influence their perception of the oversight task, our findings likely transfer best to similarly structured jobs. Below, we discuss how our findings transfer to other configurations of human oversight: synchronous and anticipatory oversight.

A key difference that makes asynchronous oversight more \textbf{\textit{tolerable}} than synchronous oversight is scalability: in our setting, no human could meaningfully review all 2493 interactions in real time during the lecture period. Asynchronous oversight allowed them to \textit{skip} most interactions and select those they could meaningfully review in the allotted time.
Due to the scaling required by increasingly agentic and autonomous AI systems, we expect asynchronous and anticipatory oversight tasks to become more common than synchronous oversight~\cite{gaube_keeping_2026, zhu_designing_2026}. 
There, scalability can be ensured for example through high-level interventions on groups of overseen instances, or by shaping the space of permissible system behaviour in advance anticipatory oversight.
Still, given a manageable task load, the demands do not feel worth the effort if the oversight is not perceived as meaningful. This connects tolerability to stimulation: stimulating tasks tend to feel more \textbf{\textit{tolerable}}, likely across all configurations. 

Further, asynchronous and synchronous oversight share a similar challenge of \textbf{\textit{stimulation}}. 
In synchronous oversight, largely uneventful passive monitoring lets attention drift~\cite{endsley2003designing, parker_grote_2020, cummings2006automation}---a threat to \textit{stimulation} of ``good'' oversight work~\cite{faas_design_2026} and the \textit{self-control} that effective oversight presupposes~\cite{sterz_quest_2024, gaube_keeping_2026}. 
Asynchronous oversight trades one form of this liability for another: at scale, it is less uneventful but more repetitive. The mechanism differs, but the consequence---little \textit{stimulation} and \textit{self-control}---does not. This challenge might not apply to anticipatory oversight, given that identifying and specifying constraints is plausibly more varied work than repetitive or uneventful monitoring. However, this form of oversight has not been researched empirically yet. 

Asynchronous and anticipatory oversight also decouple oversight from ongoing interactions, avoiding the time-critical pressure that limits timing \textbf{\textit{autonomy}} in synchronous settings. Method autonomy transfers better across configurations, depending on the intervention type: synchronous oversight permits direct control over the AI output (e.g., withholding or overriding it) alongside the indirect methods available asynchronously and anticipatorily. However, this indirect, temporally decoupled control makes the impact of interventions less clear and visible---so we expect \textbf{\textit{mastery}} to be a greater concern for indirect interventions. 
Related to these forms of interventions, prominent accounts of effective oversight require causal power~\cite{sterz_quest_2024, baum_constitutive_2026, gaube_keeping_2026}, which was reduced to one \AIbehavior\ in our deployment. When indirect interventions risk \textit{mastery} through unclear impact, tracing causal power might be equally obscured, which complicates judgments about effectiveness.

Finally, low \textbf{\textit{relatedness}} appears not to be unique to asynchronous oversight. On the one hand, we saw opportunities to utilise oversight insights for other relational work tasks. On the other hand, we found that relational oversight may be less crucial for jobs where other work tasks already fulfil relational needs than for jobs where oversight is a worker's main task. Synchronous oversight can offer more direct and more visible impact on the people affected, but little more opportunity for connecting with them. Anticipatory oversight offers clearer visibility into effects on the AI itself, but leaves the impact on people similarly invisible as asynchronous oversight. Across all forms, supporting overseers' ability to connect with others remains an important open challenge that requires deliberate attention.

\subsection{Limitations \& Future Work}
\label{subsec:limitations}
Closely linked to our action research approach, we could only accompany eight participants, limited by the number of tutors in the institution's lecture and our aim to observe a realistic everyday work task. While more participants would likely increase heterogeneity, provide further nuance, and more consistent quantitative results that could be analysed statistically, our sample sufficed to identify which characteristics troubled tutors most, along with the specific challenges and potential improvements tied to them.
Similarly, the three-month lecture period constrained us: after the initial observation phase, participants had only about 3 weeks with the modified interface. Further observations and repeated action research cycles would have offered more nuanced insights into long-term effects.
Importantly, the oversight task is largely influenced by the work-design level. If the lecture team changed the amount of weekly required reviews, likely the oversight's work characteristics---especially tolerability---would change accordingly.
Further, our setting makes a tension visible that the oversight literature has only begun to name~\cite{gaube_keeping_2026}: overseeing a conversational AI system that interacts directly with people means reading what those people write to it. Designing oversight work to be more relational increases rather than reduces the visibility of individual students to individual tutors. While our \studentFeedback\ feature was the clearest case it can also replace undesired workarounds with governed infrastructure. The relational needs of overseers and the privacy interests of the overseen are not automatically aligned, and interface work in this space should treat that trade-off explicitly.
Finally, our action research approach aims at the transferability of our findings to oversight tasks with similar characteristics, as discussed in \autoref{subsec:discussionOversight}, rather than being broadly generalisable.

These limitations underscore the need for further research across varied oversight forms and job contexts, and for more targeted, larger-scale, longitudinal studies of oversight's work characteristics and design.
While the effects of our UI modifications are likely tied to the specific task and tutors studied, we expect challenging work characteristics to transfer to other overseers and structurally similar tasks: the same design goals may inform task-specific UI solutions elsewhere.
\section{Conclusion}
While human oversight tasks can lack ``good'' work characteristics, their main challenges vary by task. Using an action research approach, we investigated the work characteristics of an oversight task performed by tutors of a statistics lecture in a longitudinal field study. 
Together with the tutors, we identified \textit{stimulation}, \textit{mastery}, and \textit{relatedness} as their primary concerns and implemented UI modifications to address these characteristics. 
Our results suggest that work characteristics of human oversight tasks can be improved not only through task or work-design adjustments, but also through interface and interaction design. We further discuss the opportunities and limitations interaction design brings for ``good'' oversight work and their transferability to other oversight tasks. 
This paper contributes to the human oversight literature by documenting how the work characteristics of a real oversight task develop over an extended period, by showing that interface design can improve some of them---most clearly stimulation and peer connection---and by identifying challenges of asynchronous oversight that interaction design cannot solve.

\begin{acks}
The work of Cedric Faas, Markus Langer, Kevin Baum, and Anna Maria Feit was partially funded by the Deutsche Forschungsgemeinschaft (DFG, German Research Foundation) -- Project-ID 389792660 -- TRR~248 (CPEC, \url{https://perspicuous-computing.science}). The work of Felix Jahn, Julius Gabelmann, and Kevin Baum was further supported by the German Federal Ministry of Education and Research (BMBF) and its successor, the Federal Ministry of Research, Technology and Space (BMFTR), within the projects MAC-MERLin (grant no.~16IW24007) and SAgA (grant no.~16IW26005). The work of Lisa Dargasz, Julius Gabelmann, and Kevin Baum was further supported by the European Regional Development Fund (ERDF) and the Saarland within the project ToCERTAIN (ID: EFRE-AuF-0000942).
\end{acks}

\bibliographystyle{ACM-Reference-Format}
\bibliography{references}

\appendix
\newpage
\section{Questionnaire Items}
\label{app:QuestionnaireItems}

\subsection{BPNS}
Items measured Autonomy (1, 4(R), 8, 11(R), 14, 17, 20(R)), Competence (3(R), 5, 10, 13, 15(R), 19(R)), and Relatedness (2, 6, 7(R), 9, 12, 16(R), 18(R), 21) with reversed scores indicated by (R).

\begin{enumerate}
    \item I feel like I am free to decide for myself how to live my life.
    \item I really like the people I interact with.
    \item Often, I do not feel very competent. (R)
    \item I feel pressured in my life. (R)
    \item People I know tell me I am good at what I do.
    \item I get along with people I come into contact with.
    \item I pretty much keep to myself and don't have a lot of social contacts. (R)
    \item I generally feel free to express my ideas and opinions.
    \item I consider the people I regularly interact with to be my friends.
    \item I have been able to learn interesting new skills recently.
    \item In my daily life, I frequently have to do what I am told. (R)
    \item People in my life care about me.
    \item Most days I feel a sense of accomplishment from what I do.
    \item People I interact with on a daily basis tend to take my feelings into consideration.
    \item In my life I do not get much of a chance to show how capable I am. (R)
    \item There are not many people that I am close to. (R)
    \item I feel like I can pretty much be myself in my daily situations.
    \item The people I interact with regularly do not seem to like me much. (R)
    \item I often do not feel very capable. (R)
    \item There is not much opportunity for me to decide for myself how to do things in my daily life. (R)
    \item People are generally pretty friendly towards me.
\end{enumerate}

\subsection{WDQ}
Insert either \textit{tutoring job} or \textit{oversight task} for X. Items were adapted to fit our study task and measure: Stimulating (9-11), Mastery (6-8), Autonomous (1-3), Relational (4,5), and Tolerable (12-14; reverse coded) 
\begin{enumerate}
    \item The X allows me to decide on the order in which things are done on the job.
    \item The X allows me to make a lot of decisions on my own.
    \item The X gives me considerable opportunity for independence and freedom in how I do the work.
    \item The results of my X are likely to significantly affect the lives of others.
    \item I have the chance in my X to get to know others.
    \item The X itself provides me with information about my performance.
    \item I receive feedback on my performance in the X from others in the team (such as the professor, teaching assistant, and other tutors).
    \item I have clear planned goals and objectives for my X.
    \item The X requires that I engage in a large amount of thinking.
    \item The X involves solving problems that have no obvious correct answer.
    \item The X requires a variety of skills.
    \item I am under constant pressure at my X. (R)
    \item I do things as part of my X which are accepted by one, but not by another. (R)
    \item The demands of my X interfere with my home and family life. (R)
\end{enumerate}

\subsection{SUS} 
\begin{enumerate}
    \item I think that I would like to use this system frequently.
    \item I found the system unnecessarily complex.
    \item I thought the system was easy to use.
    \item I think that I would need the support of a technical person to be able to use this system.
    \item I found the various functions in this system were well integrated.
    \item I thought there was too much inconsistency in this system.
    \item I would imagine that most people would learn to use this system very quickly.
    \item I found the system very cumbersome to use.
    \item I felt very confident using the system.
    \item I needed to learn a lot of things before I could get going with this system.
\end{enumerate}

\newpage
\section{Affinity Diagrams}
\label{app:affinityDiagrams}
\begin{figure}[ht]
    \centering
    \begin{subfigure}{.45\linewidth}
        \centering
    \includegraphics[page=1,width=\linewidth]{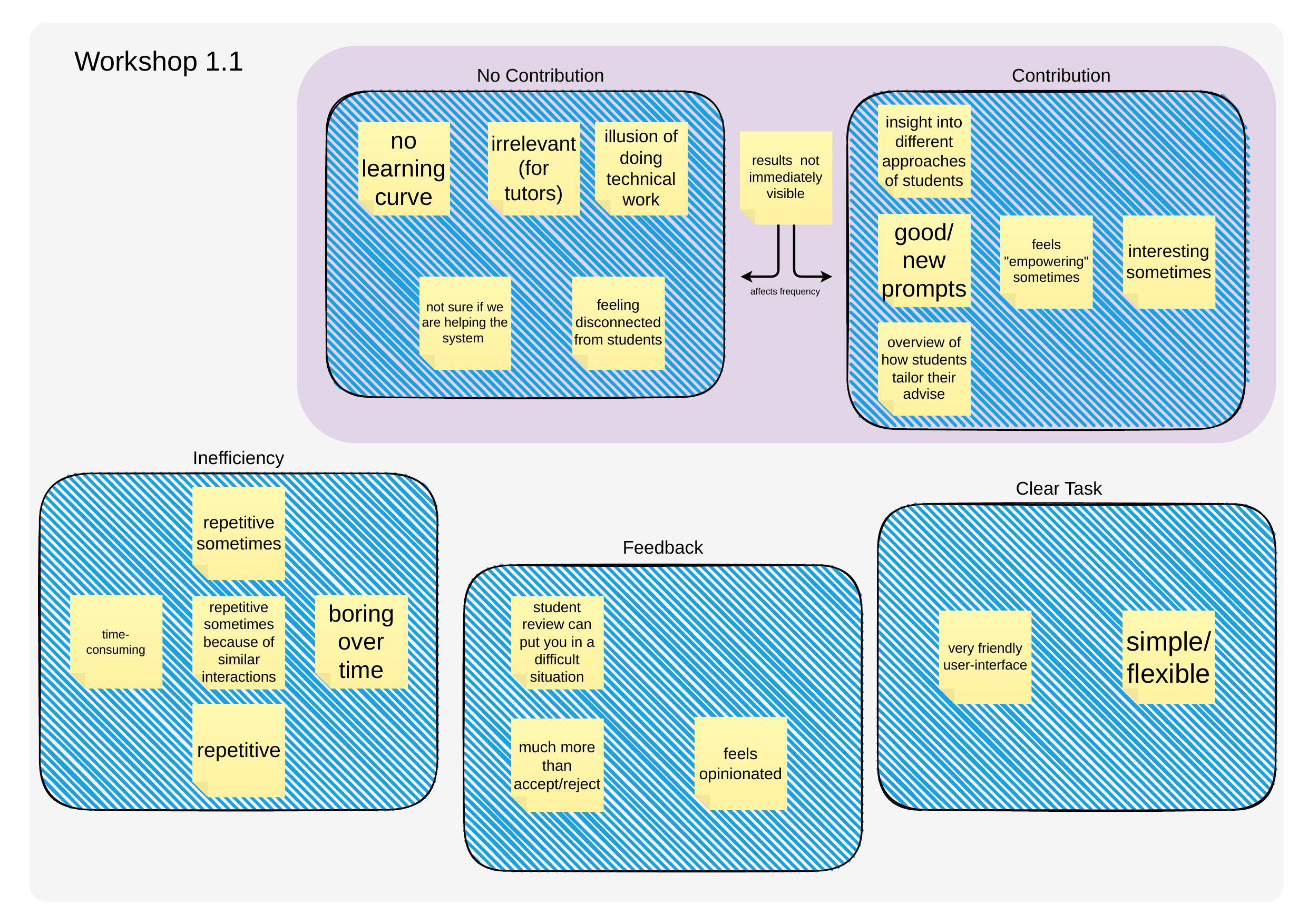}
        \caption{Workshop 1 Group 1}
    \end{subfigure}
    \begin{subfigure}{.45\linewidth}
        \centering
    \includegraphics[page=2,width=\linewidth]{images/Affinity_Diagrams.pdf}
        \caption{Workshop 1 Group 2}
    \end{subfigure}
    \begin{subfigure}{.45\linewidth}
        \centering
        \includegraphics[page=3,width=\linewidth]{images/Affinity_Diagrams.pdf}
        \caption{Workshop 2 Group 1}
    \end{subfigure}
    \begin{subfigure}{.45\linewidth}
        \centering
        \includegraphics[page=4,width=\linewidth]{images/Affinity_Diagrams.pdf}
        \caption{Workshop 2 Group 2}
    \end{subfigure}
    \caption{Affinity Diagrams created in the Workshops}
    \label{fig:affinity diagrams}
    \Description{
    This figure shows the affinity diagrams created in the workshops. A PDF version of the diagrams is published on Zenodo.
    }
\end{figure}

See \autoref{fig:affinity diagrams} for the digitized affinity diagrams that resulted from the individual workshop groups.

\newpage
\section{Work Design Questionnaire Oversight}
\label{app:wdq}

\begin{figure}[ht]
    \centering
    \includegraphics[width=\linewidth]{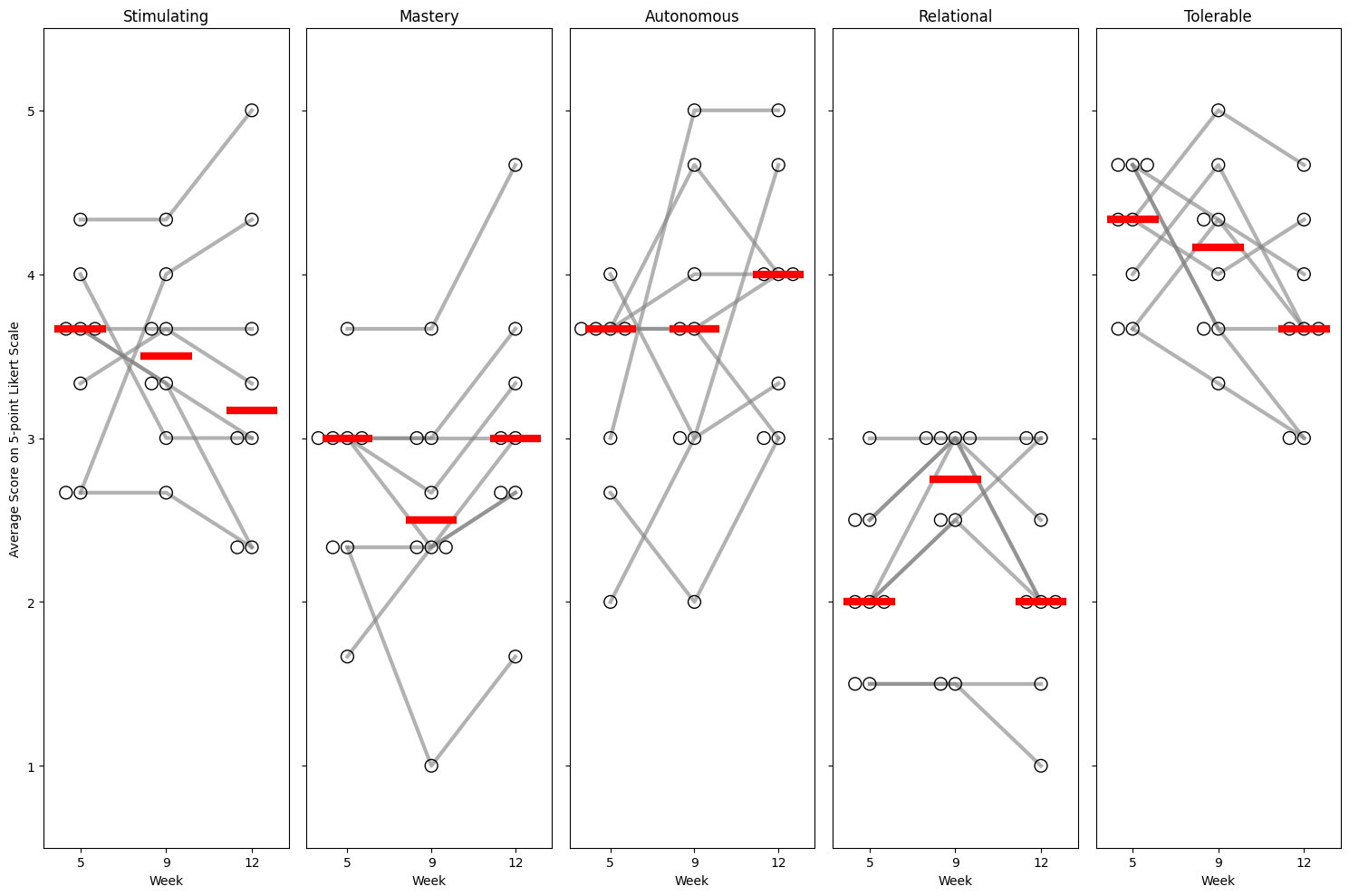}
    \caption{Work Characteristics of the Oversight Task on a Five-Point Likert Scale.}
    \label{fig:wdq-o}
    \Description{
    Plot of participants' ratings of stimulating, mastery, autonomous, relational, and tolerable. 
    Median stimulating was initially 3.7 (ranging from 2.5 to 4.5), decreased in week 9 to 3.5 (ranging from 2.5 to 4.5), and further decreased in week 12 to 3.2 (ranging between 2 and 5). 
    Median mastery was initially 3 (ranging between 1.5 and 4), decreased in week 9 to 2.5 (ranging between 1 and 4), and increased again in week 12 to 3 (ranging between 1.5 and 5). 
    Median autonomous was initially 3.6 (ranging from 2 to 4), remained in week 9 at 3.6 (ranging from 2 to 5), and increased again in week 12 to 4 (ranging from 3 to 5). 
    Median relational was initially 2 (ranging from 1.5 to 3), increased in week 9 to 2.8 (ranging from 1.5 to 3), and decreased again in week 12 to 2 (ranging from 1 to 3). 
    Median tolerable was initially 4.3 (ranging between 3.5 and 5), decreased in week 9 to 4.2 (ranging between 3 and 5), and further decreased in week 12 to 3.7 (ranging between 3 and 5).
    }
\end{figure}

In \autoref{fig:wdq-o}, participant's responses to the repeated questionnaires about the oversight task's work characteristics are plotted. The median score for each characteristic in each week is indicated by the thick red line.

\newpage
\section{Need Fulfillment}
\label{app:needs}
\begin{figure}[ht]
    \centering
    \includegraphics[width=\linewidth]{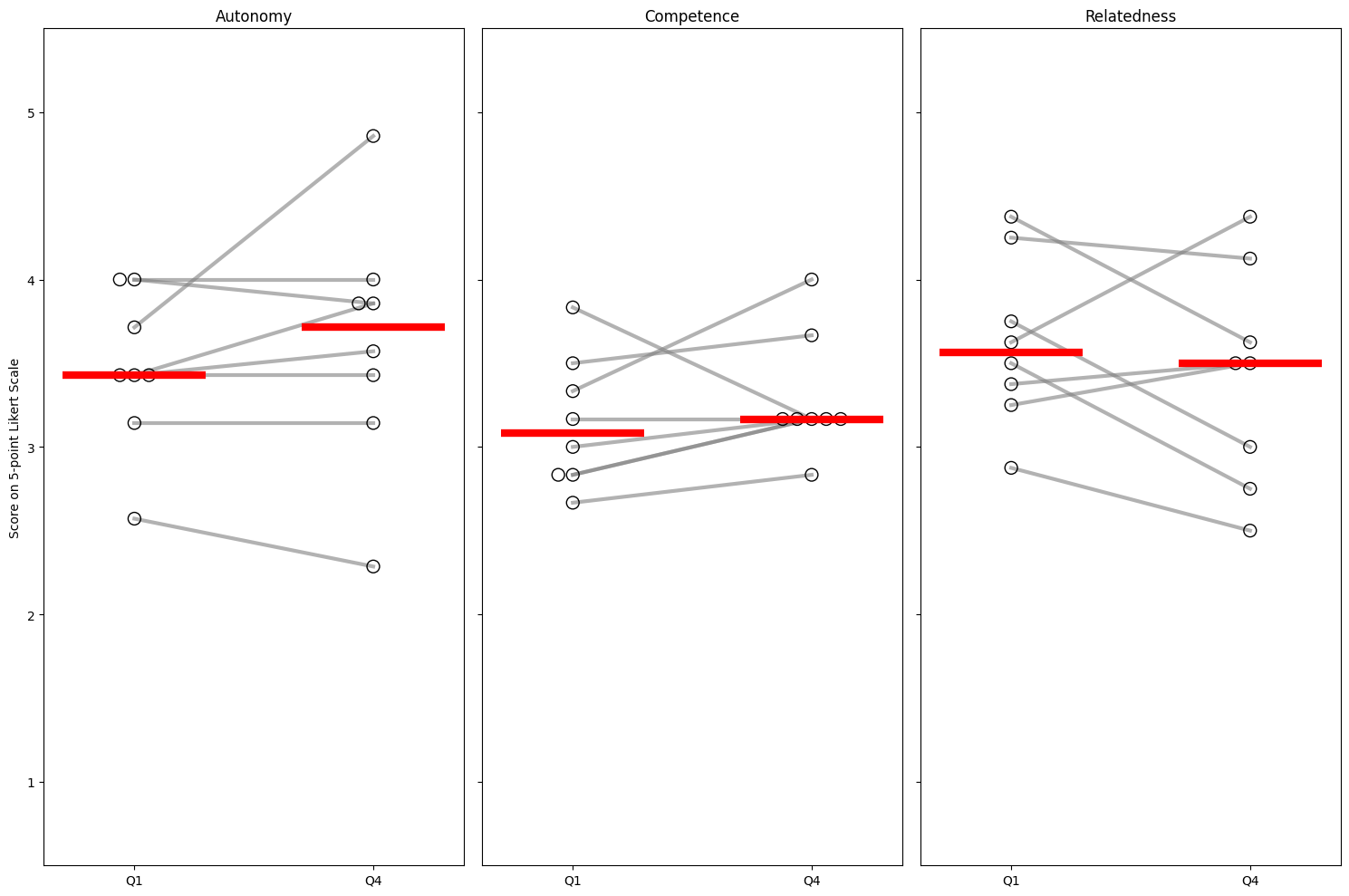}
    \caption{Participants' Need Fulfillment on a Five-Point Likert Scale.}
    \label{fig:needs}
    \Description{
    Plot of participants' ratings of autonomy, competence, and relatedness. Median autonomy was initially 3.4 (ranging from 2.5-5) and increased to 3.7 (ranging between 2 and 5) after the UI modification. Median Competence was initially 3.1 and increased to 3.2 after the UI modification. For competence, all ratings ranged between 2.5 and 4. Median relatedness was initially 3.6 and decreased to 3.5 after the UI modification. For relatedness, all ratings ranged between 2.5 and 4.5.
    }
\end{figure}

In the first and last questionnaire, we used questionnaire items to measure the needs of \textit{Autonomy}, \textit{Competence}, and \textit{Relatedness}~\cite{ryan_self-determination_2018, ryan_self-determination_2000, gagne_role_2003}. There were no clear changes between the two assessments observable (\autoref{fig:needs}).

\newpage
\section{Oversight Interface Usability}
\label{app:sus}
\begin{figure}[ht]
    \centering
    \includegraphics[width=\linewidth]{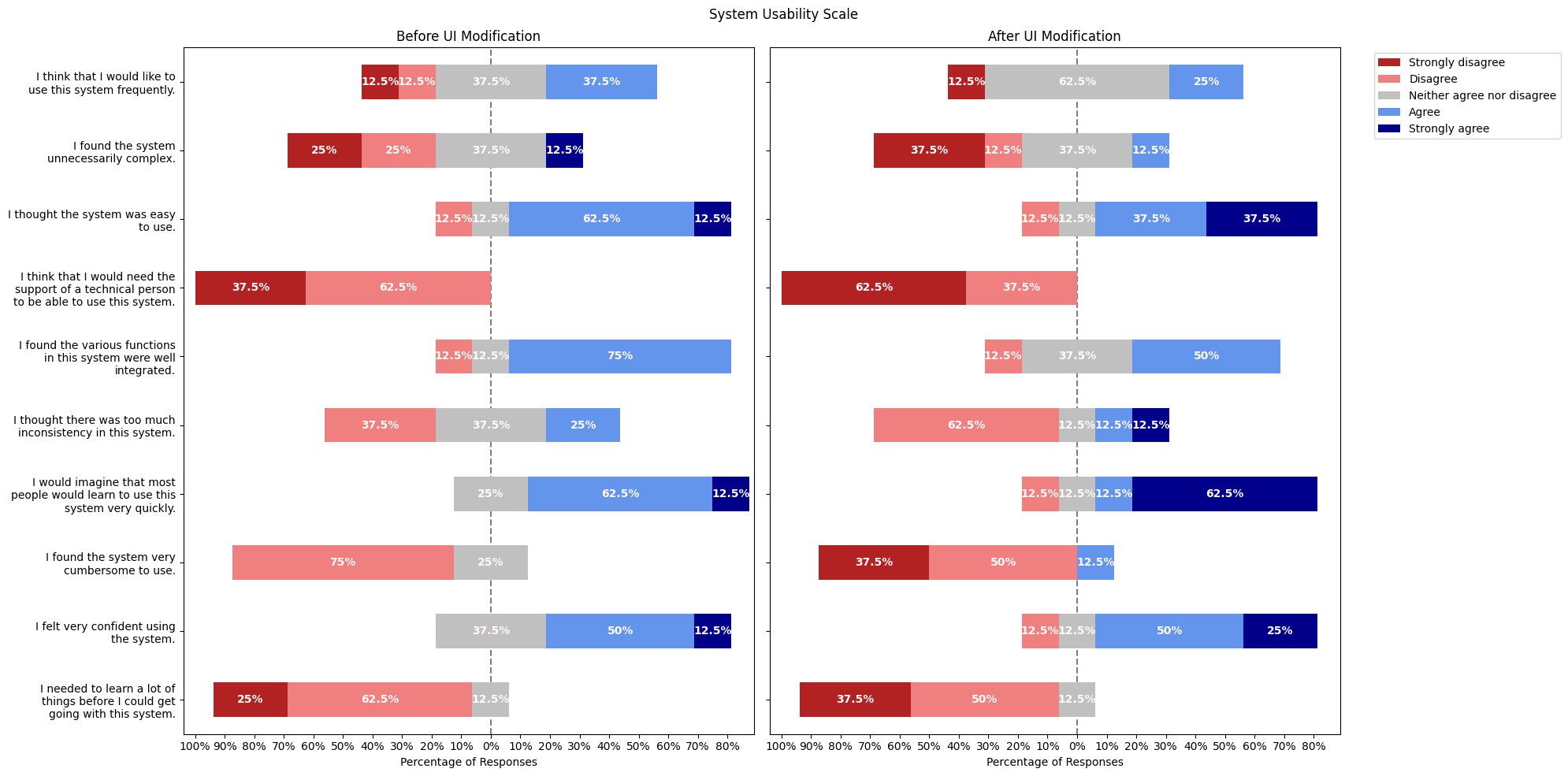}
    \caption{System's Usability on a Five-Point Likert Scale.}
    \label{fig:sus}
    \Description{Stacked bar chart of participants' ratings for the system's usability, comparing it before the UI modifications with after. The usability of the interface was slightly improved for most items in the scale. Ratings of inconsistencies in the system, the UI being cumbersome to use, and participants' confidence in using it showed greater variance.}
\end{figure}

To observe whether the UI modifications changed the \textit{Usability}~\cite{brooke_sus_1996} of the oversight interface, we measured the oversight interface's usability before and after the UI modifications. Participants' answers did show a tendency to a better usability of the modified interface, with some polarising perceptions about inconsistencies, cumbersome use, and users' confidence (\autoref{fig:sus}).

\end{document}